\documentclass[pdflatex,sn-mathphys-num]{sn-jnl}
\usepackage{multirow}%
\usepackage{amsmath,amssymb,amsfonts}%
\usepackage{amsthm}%
\usepackage{mathrsfs}%
\usepackage[title]{appendix}%
\usepackage{xcolor}%
\usepackage{textcomp}%
\usepackage{manyfoot}%
\usepackage{booktabs}%
\usepackage{algorithm}%
\usepackage{algorithmicx}%
\usepackage{algpseudocode}%
\usepackage{listings}%
\usepackage{tabularx}

\usepackage{url}
\usepackage{makecell}
\usepackage{hyperref}

\usepackage{rotating}
\usepackage{graphicx}
\graphicspath{ {./}, {./plots/} }
\DeclareGraphicsExtensions{ .pdf, .png}

\usepackage{lineno}

\usepackage{xspace}

\newcommand{\snn}{\sqrt{s_{_{\textsc{nn}}}}}

\newcommand{\ampt}{{\textsc{ampt}}}
\newcommand{\hijing}{{\textsc{hijing}}}

\newcommand{\pt}{p_{T}}
\newcommand{\gevc}{GeV/$c$}
\newcommand{\Nch}{N_{\rm ch}}
\newcommand{\obs}{{\text{obs}}}
\newcommand{\cme}{{\textsc{cme}}}
\newcommand{\bkg}{{\text{bkg}}}
\newcommand{\cl}{{\text{cl}}}
\newcommand{\sms }{\scriptstyle}
\newcommand{\Ru}{\rm Ru+Ru}
\newcommand{\Zr}{\rm Zr+Zr}

\newcommand{\dg}{\Delta\gamma}
\newcommand{\dgv}{(\Delta\gamma/v_2)}
\newcommand{\dd}{\Delta\delta}
\newcommand{\fcme}{f_{\textsc{cme}}}
\newcommand{\minv}{m_{\rm inv}}

\newcommand{\deta}{\Delta\eta}
\newcommand{\dphi}{\Delta\phi}
\newcommand{\srp}{{\textsc{rp}}}
\newcommand{\ssp}{{\textsc{sp}}}
\newcommand{\spp}{{\textsc{pp}}}
\newcommand{\pep}{{\textsc{ep}}}
\newcommand{\sos}{{\textsc{os}}}
\newcommand{\sss}{{\textsc{ss}}}

\newcommand{\gos}{{\ensuremath{\gamma_\sos}\xspace}}
\newcommand{\gss}{{\ensuremath{\gamma_\sss}\xspace}}
\newcommand{\gab}{{\gamma_{\alpha\beta}}}
\newcommand{\two}{{\{2\}}}
\newcommand{\EP}{\{\pep\}}

\newcommand{\zdc}{\{{\textsc{zdc}}\}}
\newcommand{\tpc}{\{{\textsc{tpc}}\}}

\newcommand{\vtwopp}{v_2^\spp}
\newcommand{\vtwosp}{v_2^\ssp}
\newcommand{\dgpp}{\dg^\spp}
\newcommand{\dgsp}{\dg^\ssp}

\newcommand{\psiep}{\ensuremath{\psi_{\textsc{EP}}\xspace}}
\newcommand{\psirp}{\ensuremath{\psi_{\textsc{RP}}\xspace}}

\newcommand {\mean}[1]   {\langle{#1}\rangle}

\begin{document}

\title{Experimental review on the chiral magnetic effect in relativistic heavy ion collisions}

\author[1]{\fnm{Wei} \sur{Li}}\email{wl33@rice.edu}

\author[2]{\fnm{Qiye} \sur{Shou}}\email{shouqiye@fudan.edu.cn}

\author[3]{\fnm{Fuqiang} \sur{Wang}}\email{fqwang@purdue.edu}

\affil[1]{\orgdiv{Physics and Astronomy Department}, \orgname{Rice University}, \orgaddress{\city{Houston}, \postcode{77005}, \state{Texas}, \country{USA}}}

\affil[2]{\orgdiv{Key Laboratory of Nuclear Physics and Ion-beam Application (MOE), Institute of Modern Physics}, \orgname{Fudan University}, \orgaddress{\city{Shanghai}, \postcode{200433}, \country{China}}}

\affil[3]{\orgdiv{Department of Physics and Astronomy}, \orgname{Purdue University}, \orgaddress{\city{West Lafayette}, \postcode{47907}, \state{Indiana}, \country{USA}}}

%
%


\abstract{The chiral magnetic effect (CME) refers to a predicted phenomenon in quantum chromodynamics that manifests as a charge separation along an external magnetic field, driven by an imbalance of quark chirality. Searches for the CME have been carried out by azimuthal particle correlations in relativistic heavy ion collisions where such a chirality imbalance is anticipated and a strong magnetic field is created in the initial stage. No conclusive experimental evidence on the CME has been established so far because of large background contributions to azimuthal correlation observables. We review the status of the experimental search for the CME, covering the observables used, the techniques to mitigate backgrounds, and the strengths and limitations of various experimental approaches, and outline a future prospect of the CME search in high-energy nuclear collisions.}


\maketitle
\section{Introduction}
A primary goal of the heavy ion physics programs at the Relativistic Heavy Ion Collider (RHIC) and the Large Hadron Collider (LHC) is to create and study the quark-gluon plasma (QGP), a state predicted by Quantum Chromodynamics (QCD) to exist at high temperatures and/or baryon densities, in which quarks and gluons are deconfined over extended volumes much larger than that of a hadron~\cite{Shuryak:2008eq}. 
In this deconfined regime, chiral symmetry, spontaneously broken in the QCD vacuum, is expected to be approximately restored, allowing quarks to behave as nearly massless, chiral particles.
An intriguing property of high-temperature QCD is vacuum fluctuations into topological states where gluon fields possess nonzero Chern-Simons winding numbers. Interactions of quarks with such gluon fields will produce metastable domains of chirality imbalance with unequal numbers of left- and right-handed quarks~\cite{Kharzeev:1998kz}. Such a chirality imbalance, under a strong magnetic field, can generate an electric current, or a charge separation along the direction of the magnetic field. This phenomenon is known as the chiral magnetic effect (CME)~\cite{Kharzeev:2007jp,Fukushima:2008xe}. The CME manifests violations of local parity and charge-parity symmetries in the strong interaction and is closely connected to the QCD axial anomaly, a key ingredient in understanding chiral symmetry breaking and mass generation in QCD~\cite{tHooft:1986ooh}. 

Relativistic heavy ion collisions provide an ideal platform to search for the CME.
The high temperatures achieved in these collisions can significantly enhance the vacuum transition probabilities~\cite{Kharzeev:1999cz,Kharzeev:2004ey,Kharzeev:2007jp}.  
A strong magnetic field on the order of $B\sim m_\pi^2/e \sim 10^{18}$~Gauss is believed to be created in non-central heavy ion collisions (here $e$ is the elementary charge and $m_\pi$ is the pion mass)~\cite{Skokov:2009qp,Deng:2012pc,Voronyuk:2011jd}.  
This magnetic field is expected to quickly fade off as the two spectator remnants recede from each other. 
However, the decay time can be significantly prolonged depending on the electric conductivity of the QGP~\cite{Tuchin:2013apa,McLerran:2013hla,Kharzeev:2009pj,Li:2018ufq,Huang:2022qdn}.
With plausible parameters for the axial current density and the time evolution of the magnetic field, anomalous viscous fluid dynamics (\textsc{avfd}) calculations suggest that an observable CME signal could indeed manifest itself in relativistic heavy ion collisions~\cite{Yin:2015fca,Shi:2017cpu,Jiang:2016wve}.
On the other hand, a significantly smaller signal than experimentally perceptible has also been theoretically suggested~\cite{Muller:2010jd}.

Experimental searches for the CME in relativistic heavy ion collisions at RHIC and the LHC have persisted over the last two decades. 
Significant progress has been made, while many challenges related to the treatment of background effects remain open. 
Here, we review the experimental status of the CME search and discuss prospects of a potential CME discovery in forthcoming high-statistics heavy ion data sets.
For further reading, the reader is referred to extensive reviews on the subject in Refs.~\cite{Kharzeev:2015znc,Zhao:2018ixy,Zhao:2019hta,Li:2020dwr,Qi-Ye:2023zyf,Kharzeev:2024zzm,chen_properties_2024,shou_properties_2024,Feng:2025yte}.  
We note that this is an experimental review so the theoretical description/discussion is kept  minimal, mostly at the level needed to interpret observables. For in-depth theoretical discussions, the reader is referred to extensive literature on the subject, for example, Refs.~\cite{Kharzeev:2009fn,Kharzeev:2013ffa,Huang:2015oca,Hattori:2016emy,Kharzeev:2020jxw,Kharzeev:2015kna}.

The rest of the review is organized as follows. Section~\ref{sec:method} describes the main observable used in most of the CME searches, the physics backgrounds associated with the observable, and experimental ways to mitigate these backgrounds. Section~\ref{sec:result} presents early measurements, discusses the physics backgrounds in these measurements, and highlights the recent measurements where these backgrounds are mitigated. Finally, a summary and outlook is given in Sect.~\ref{sec:summary}. 

\section{Experimental observables and analysis methods}\label{sec:method}

\subsection{The $\dg$ correlator}
The CME yields an emission pattern of positively charged particles in one direction and negatively charged particles in the opposite direction along the magnetic field event-by-event. Because the magnetic field is on average perpendicular to the reaction plane (RP), the CME signal can be expressed as a sine term in particle's azimuthal angle ($\phi$) relative to the RP's ($\psirp$) of the Fourier expansion of particle distributions~\cite{Voloshin:1994mz},
\begin{equation}
  \label{eq:Fourier}
  \frac{dN_\pm}{d\phi} \propto 1 + 2v_1\cos(\phi-\psirp)
  + 2v_2\cos2(\phi-\psirp) + \cdots
    + 2a_{1\pm}\sin(\phi_\pm-\psirp) + \cdots \,.
\end{equation}
The cosine terms are the usual flow harmonics, a result of hydrodynamic-type expansion from an initial anisotropic overlap region of heavy ion collisions~\cite{Ollitrault:1992bk}; 
the parameter $v_1$ is often called directed flow, and $v_2$ elliptic flow. 
The subscripts $\pm$ denote the electric charge sign of the particle.  
Neglecting effects of the electromagnetic interaction, the $v_n$ coefficients are charge independent.  

The coefficients $a_{1+} = -a_{1-}$ in Eq.~(\ref{eq:Fourier}) quantify the parity ($\mathcal{P}$)-odd charge asymmetry associated with the CME.
The measurements of $\mean{a_{1\pm}}$ over many events, however, vanish because the sign of the topological charge fluctuates randomly from event to event. One can only measure the charge-dependent correlations such as $\mean{a_{1+}a_{1-}}=-a_1^2$ and $\mean{a_{1\pm}a_{1\pm}}=a_1^2$ that would remain finite in the presence of CME.

\begin{figure}[hbt]
  \begin{minipage}{0.5\textwidth}
    \includegraphics[width=\textwidth]{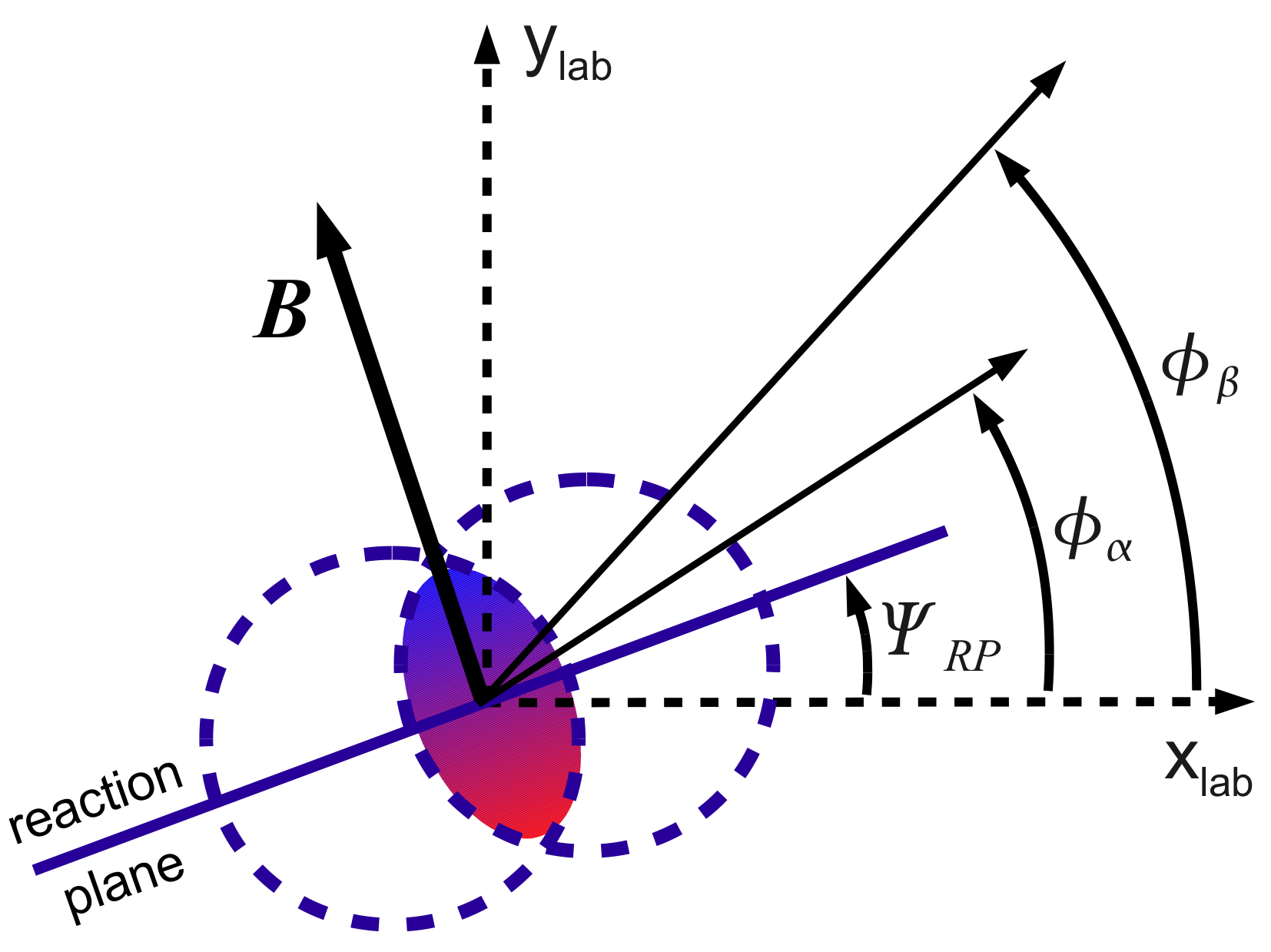}
  \end{minipage}\hfill
  \begin{minipage}{0.45\textwidth}
    \vspace*{1.5cm}
    \caption{(Color online) Schematic view of the transverse plane in a collision of two heavy nuclei -- one emerging from and one going into the page. The azimuthal angles of the reaction plane and produced
      particles with charges $\alpha$ and $\beta$ as used in Eqs.~(\ref{eq:gamma}) and (\ref{eq:c3}) are depicted here. 
      Drawing is taken from Ref.~\cite{STAR:2009wot}.}
    \label{fig:planes}
  \end{minipage}
\end{figure}

A commonly used observable sensitive to these correlations is the charge-dependent three-point azimuthal correlator~\cite{Voloshin:2004vk},
\begin{equation}
  \gamma_{\alpha\beta} \equiv \mean{ \cos(\phi_\alpha +\phi_\beta -2\psirp)}\,,
  \label{eq:gamma}
\end{equation}
where $\phi_\alpha$ and $\phi_\beta$ are the azimuthal angles of two particles of interest (POI) with charge signs $\alpha$ and $\beta$, $\psirp$ is the azimuthal angle of the RP defined by the beam direction and the impact parameter vector.
The averaging is taken over all particle pairs in an event and over all events.
Figure~\ref{fig:planes} illustrates the geometry of a non-central heavy ion collision and the relevant azimuthal angles.

The RP and the impact parameter are theoretical concepts and can only be rigorously defined under the limit of smooth nuclear density distributions. Because nucleon distributions in nuclei are lumpy, the symmetry planes relevant for collective flow are determined by the geometry of the participating nucleons~\cite{Alver:2008zza}. They are also called participant planes (PP) or flow planes, and they fluctuate event-by-event, differing for each harmonic order. 
Because the magnetic field is determined mostly by spectator protons, its direction is expected to be roughly perpendicular to the first-order symmetry plane, which is often called the spectator plane (SP).
Thus, the $\psirp$ in the various terms of Eq.~(\ref{eq:Fourier}) should be replaced by the corresponding-order harmonic planes ($\psi_n$). 
In particular, that for the CME sine term can be replaced by the first-order symmetry plane, which is most closely correlated to the RP. 
It can also be replaced by the second-order PP angle ($\psi_2^\spp$) because it is closely related to the RP and the most accurately assessed in experiment; in this case, the coefficients $a_\pm$ would be reduced accordingly.

The geometry-defined symmetry planes cannot be experimentally measured. They leave imprint in and thus can be estimated from  azimuthal distributions of the final-state particles~\cite{Poskanzer:1998yz},  
\begin{equation}\label{eq:psi}
  \psi_{n,\pep}
  = \frac{1}{n} \arctan\left(
  \frac{\mean{\sin n \phi}}{\mean{\cos n \phi}} \right)\,,
\end{equation}
where the average is taken over all particles in an event in a given momentum region.  
The experimentally reconstructed $\psi_{n,\pep}$, called the event plane (EP), deviates from the geometry-defined $\psi_n$ due to the finite number of particles used in its reconstruction, the effect of which can be quantified by the so-called EP resolution, $R_\pep \equiv\mean{\cos  n(\psi_{n,\pep}-\psi_n)}$. 
The latter can be evaluated from the correlations between different EPs, e.g. by the sub-event method~\cite{Poskanzer:1998yz}. 
The $\gamma$ correlator measured using the reconstructed $\psi_{n,\pep}$, in place of the theoretical $\psirp$ in Eq.~(\ref{eq:gamma}), is then
\begin{equation}
  \gamma_{\alpha\beta} = \mean{ \cos(\phi_\alpha +\phi_\beta -2\psi_{2,\pep})}/R_\pep\,.
  \label{eq:gammaEP}
\end{equation}

Similar to Eq.~(\ref{eq:gamma}), the $\gamma$ correlators can also be measured by correlating the particles $\alpha$ and $\beta$ to a third charge-inclusive particle $c$~\cite{Voloshin:2004vk,STAR:2009tro,STAR:2009wot},
\begin{subequations}\label{eq:c3g} 
\begin{align}
    C_{3,\alpha\beta} &= \mean{\cos(\phi_\alpha+\phi_\beta-2\phi_{c})}\,, \label{eq:c3} \\
    \gab &= C_{3,\alpha\beta} / v_{2,c}\,. \label{eq:c3v2}
\end{align}
\end{subequations}
The azimuthal direction of particle $c$ in this case plays the role of $\psirp$ in Eq.~(\ref{eq:gamma}) but with a finite resolution equal to the elliptic flow parameter $v_{2,c}$ of particle $c$.

The $v_2$ coefficient, corrected for the EP resolution, is calculated as
\begin{equation}\label{eq:v2}
    v_2\EP = \mean{\cos 2(\phi-\psi_{n,\pep})}/R_\pep\,.
\end{equation}
Similarly, the measurements of $v_2$ can be done with a two-particle cumulant method as 
\begin{subequations}\label{eq:v22}
\begin{align}
    V_2 &= \mean{\cos 2(\phi_1-\phi_2)}\,, \label{eq:V2} \\
    v_2\two &\equiv \sqrt{V_2}\,,
\end{align}
\end{subequations}
where both particles (azimuthal angles $\phi_1$ and $\phi_2$) are taken from the same momentum region~\cite{Poskanzer:1998yz}.  

The $\gamma$ variable in Eqs.~(\ref{eq:gamma}) and (\ref{eq:c3g}) is parity-even and is therefore subject to many background sources. 
Charge-independent background correlations, such as those caused by global momentum conservation, can be readily removed by taking the difference between the opposite-sign (OS) and same-sign (SS) correlators~\cite{Voloshin:2004vk},
\begin{subequations}\label{eq:Delta}
\begin{align}
    \Delta C_3 &\equiv C_{3,\sos} - C_{3,\sss}\,, \label{eq:dc} \\
    \dg &\equiv \gos - \gss\,. \label{eq:dg}
\end{align}
\end{subequations}
The remaining charge-dependent background in $\dg$ represents the major problem in the CME search. 
They are discussed in detail in Sect.~\ref{sec:method:bkgd}.

Several other observables for the CME search have also been proposed~\cite{Du:2008zzb,Finch:2018ner,Ajitanand:2010rc,Magdy:2017yje,Feng:2018chm,Feng:2020cgf,Bzdak:2011np,Sun:2018onk,Tang:2019pbl,Feng:2019pxu,Li:2024pue,Li:2024pue}.  
Most of them are similar (or proven to be equivalent) to the $\gamma$ correlators. 
The $\dg$ observable has become the most widely studied observable in search for the CME. 
We will concentrate on $\dg$ in this review.

\subsection{Charge-dependent physics backgrounds}\label{sec:method:bkgd}
Because the $\gamma$ correlators, unlike the CME $a_1$ parameter in Eq.~(\ref{eq:Fourier}), are parity even and thus subject to QCD physics backgrounds. Charge-independent backgrounds cancel in the $\dg$ correlator. However, many physics processes are intrinsically charge dependent; examples are resonance decays and (mini-)jet correlations. There are more contributions to OS pairs than to SS pairs from resonance decays. Charge-ordering is well known in jet fragmentation that also leads to an excess of OS over SS correlations. These short-range charge-dependent correlations can be generally characterized as ``clusters''. Local charge conservation (LCC) is preserved in cluster decays, so these background contributions are often referred to as LCC effects. 

Although LCC correlations are local in phase space, their coupling to the collective flow can generate RP–dependent modulations that extend over large pseudorapidity separations, becoming an important background to the CME search.

\subsubsection{Reaction-plane dependent (flow-induced) backgrounds}
The major backgrounds to $\dg$ are those charge-dependent cluster correlations, or LCC effects, that are made dependent of the RP by the elliptic flow of those clusters~\cite{Voloshin:2004vk,Wang:2009kd,Bzdak:2009fc,Petersen:2010di,Pratt:2010zn,Bzdak:2010fd,Schlichting:2010qia,Liao:2010nv,Bzdak:2012ia,Toneev:2012zx}. 
It can be expressed schematically as
\begin{equation}
\label{eq:cluster}
  \dg_\bkg = \frac{N_\cl}{N_\alpha N_\beta} \cdot \mean{\cos(\phi_\alpha+\phi_\beta-2\phi_\cl)} \cdot v_{2,\cl}\,,
\end{equation}
where $v_{2,\cl}$ is the elliptic flow, $\phi_\cl$ is the emission azimuthal angle, and the POIs are two decay products of the cluster. The quantities in Eq.~(\ref{eq:cluster}) depend on the types and kinematics of clusters.

\subsubsection{Nonflow contamination}\label{sec:method:nonflow}
Collective flow is a final-state response to the initial geometry because of interactions~\cite{Ollitrault:1992bk}. It causes all particles in an event to correlate with each other. This is the foundation of the concepts of flow planes $\psi_n$ and flow harmonics $v_n$. 
However, experimentally, these quantities can only be measured/estimated by particle correlations, as in Eqs.~(\ref{eq:v2}) and (\ref{eq:v22}). 
Not all particle correlations stem from the global event-wise (flow) correlations to the collision geometry. 
Correlations not of flow origins exist, for example, those from resonance/cluster decays and jets~\cite{Borghini:2000cm,Borghini:2006yk,Wang:2008gp,Ollitrault:2009ie}. 
These correlations are generally referred to as nonflow and affect the $v_2\two$ measurement by Eq.~(\ref{eq:v22}). 
Because the EP is reconstructed from particles which contain decay products of clusters and because the $R_\pep$ is evaluated based on particle correlations~\cite{Poskanzer:1998yz}, 
nonflow affects $v_2\EP$ measurement by Eq.~(\ref{eq:v2}) as well.

Similarly, because the $\gamma$ correlators are measured by Eqs.~(\ref{eq:gammaEP}) and (\ref{eq:c3v2}), they are affected by nonflow contamination in $R_\pep$ and $v_{2,c}$. 
To mitigate these nonflow effects, it is important to construct the EP using particles that are well separated in pseudorapidity from those entering the CME-sensitive correlators. A large rapidity gap between the EP detectors and the POIs strongly suppresses short-range nonflow correlations from jets and resonance decays while preserving the long-range collective flow correlations. The nonflow effects in the CME search are  further discussed in Sects.~\ref{sec:result:isobar} and~\ref{sec:result:ppsp}.

\subsubsection{Reaction-plane independent three-particle backgrounds}
The three-particle correlator $C_{3,\alpha\beta}$ of Eq.~(\ref{eq:c3}) contains correlations not related to the RP, e.g.~from back-to-back dijets with particles $\alpha$ and $\beta$ originating from one jet and the particle $c$ from the other jet. Because the intra-jet particle correlations are charge-dependent (charge ordering), these backgrounds are present in $\dg$ and need to be taken care of by other means. 
The contribution from this RP-independent three-particle correlation background scales as an inverse of the multiplicity squared, and thus is most severe in peripheral collisions~\cite{STAR:2009tro,STAR:2009wot}.

Because the EP is reconstructed from particles, effects of the RP-independent three-particle correlations are essentially the same between calculations using the three-particle correlator of Eq.~(\ref{eq:c3}) or the EP method of Eq.~(\ref{eq:gammaEP}); the effects are easier to analyze in the former and is subtle in the latter.

We note that the terminology of ``nonflow'' was invented in contrast to flow, and thus in most cases refers to those nonflow correlations contributing to flow measurements of $v_n$.
However, more broadly, nonflow can refer to any correlations that are not of the event-wise global correlations to the initial geometry; thus, the RP-independent three-particle correlations described here is also part of nonflow. In various places of this review, we use `nonflow' to refer to both the nonflow in $v_2$ measurements and the RP-independent three-particle correlations.

\subsection{Experimental and analysis methods}
Experimentally measured CME observables can contain contributions from both the signal and various background sources. In general, one can write
\begin{equation}
\text{Observable} = \text{Signal} + \text{Background}\,.
\end{equation}
The background and signal components, although unknown {\em a priori}, generally possess different properties. Thus, a central strategy of CME searches is to vary one component while keeping the other approximately fixed, and then study the corresponding change in the observable.

There are several general approaches in dealing with backgrounds: (1) directly measuring or estimating the background strength, (2) varying the background strength to extract the signal from its dependence, (3) varying the signal strength while keeping the background approximately constant, and (4) varying both the background and signal strength in controlled and distinct ways to separate the two simultaneously. We describe these approaches in the following subsections.

\subsubsection{Mixed harmonics}\label{sec:method:kappa}
The flow-induced backgrounds are proportional to the $v_2$ values of background sources; see Eq.~(\ref{eq:cluster}). These $v_2$ values cannot be all readily measured and there are in principle countless number of these background sources. However, one may construct other azimuthal correlators whose non-CME contributions are also proportional to $v_2$, for example~\cite{Choudhury:2019ctw},
\begin{equation}
\dg_{132} \equiv \mean{\cos(\phi_\alpha-3\phi_\beta+2\psi_2)} 
\propto \mean{\cos(\phi_\alpha-3\phi_\beta+2\phi_\cl)} v_{2,\cl} \,.
\label{eq:dg132}
\end{equation}
One may also construct a doubled harmonic correlator $\dg_{224}=\mean{\cos2(\phi_\alpha+\phi_\beta-2\psi_4)}$~\cite{Voloshin:2011mx,Voloshin:2012fv}, which includes background contribution corresponding to the fourth harmonic flow. 
Although these correlators are sensitive to the flows of the background contributing sources, the sensitivity coefficients differ from that of the $\dg\equiv\dg_{112}$ correlator of Eq.~(\ref{eq:dg}).
Because the relationship of these coefficients are unknown and cannot be possibly measured for numerous background sources, these mixed-harmonic correlators cannot be quantitatively used to measure the background contribution to $\dg$.

The $\dg$ correlator benefits from the cancellation of RP-independent correlation backgrounds in $\mean{\cos(\phi_\alpha-\psirp)\cos(\phi_\beta-\psirp)}$ and $\mean{\sin(\phi_\alpha-\psirp)\sin(\phi_\beta-\psirp)}$. One may use another correlator 
\begin{equation}
    \label{eq:delta}
    \delta_{\alpha\beta} \equiv \mean{\cos(\phi_\alpha-\phi_\beta)} \,,    
\end{equation}
which is also sensitive to the CME but the RP-independent correlation backgrounds add up in $\dd\equiv\delta_\sos-\delta_\sss$ and overwhelm any CME signal.
An attempt was made to relate the background in $\dg$ to $\dd$ through an approximate relationship~\cite{Bzdak:2012ia} 
\begin{equation}
\label{eq:gamma-delta}
  \gamma_{\alpha\beta} \equiv \mean{\cos(\phi_\alpha+\phi_\beta-2\psirp))} 
  \approx \mean{\cos(\phi_\alpha-\phi_\beta)\cos2(\phi_\beta-\psirp)} 
  \approx \delta_{\alpha\beta}v_2 \,.
\end{equation}
However, the factorization in Eq.~(\ref{eq:gamma-delta}) is mathematically invalid because the angles $(\phi_\alpha-\phi_\beta)$ and $2(\phi_\beta-\psirp)$ are not independent.
The correct factorization of the terms is done in Eq.~(\ref{eq:cluster}).
The STAR Collaboration introduced a parameter, 
\begin{equation}
    \kappa_2 = \frac{\dg_{112}}{v_2\dd}\,,
    \label{eq:k2}
\end{equation}
to account for the approximations made in the steps of Eq.~(\ref{eq:gamma-delta}). 
Compared to the background contribution given in Eq.~(\ref{eq:cluster}), the $\kappa_2$ parameter is related to the azimuthal distribution of cluster decay products as
\begin{equation}
    \kappa_2 = \frac{\mean{\cos(\phi_\alpha+\phi_\beta-2\phi_{\rm cl})}}
         {\mean{\cos(\phi_\alpha-\phi_\beta)}} \cdot \frac{v_{2,{\rm cl}}}{v_2}\,.
    \label{eq:k2phys}
\end{equation}
However, the $\kappa_2$ parameter cannot be theoretically calculated or experimentally measured in a background-only scenario. Consequently, adopting {\em ad hoc} values of $\kappa_2$ leads to large systematic uncertainties in any extracted CME signal.

A related mixed-harmonic correlator~\cite{CMS:2017pah,ALICE:2020siw},
\begin{equation}
\dg_{123} \equiv \mean{\cos(\phi_\alpha+2\phi_\beta-3\psi_3)}
\propto \mean{\cos(\phi_\alpha+2\phi_\beta-3\phi_\cl)} v_{3,\cl} \,.
\label{eq:dg123}
\end{equation}
can be constructed with respect to the third-order harmonic plane and therefore contains only background contributions, since the CME does not couple to $\psi_3$.
The entire correlator is sensitive to the triangular flow $v_3$ from the cluster correlation $\mean{\cos(\phi_\alpha+2\phi_\beta-3\phi_\cl)}$. 
A similar parameter can be written as
\begin{eqnarray}
    \kappa_3 = \frac{\dg_{123}}{v_3\dd}\,,
    \label{eq:k3}
\end{eqnarray}
which in the cluster picture corresponds to
\begin{equation}
    \kappa_3 = \frac{\mean{\cos(\phi_\alpha+2\phi_\beta-3\phi_{\rm cl})}}
         {\mean{\cos(\phi_\alpha-\phi_\beta)}} \cdot \frac{v_{3,{\rm cl}}}{v_3}\,.
    \label{eq:k3phys}
\end{equation}
While the $\kappa_2$ and $\kappa_3$ parameters depend on the detailed azimuthal structure of cluster correlations and cannot be assumed identical as {\em a priori}, the CMS collaboration~\cite{CMS:2017pah} has shown that under certain conditions, when the same particle kinematics and acceptance are used for constructing all correlators, the two parameters are approximately equal, i.e.\ $\kappa_2 \simeq \kappa_3$.
Conversely, unlike $\dg_{112}$ and $\dg_{123}$, the sensitivity of $\dg_{132}$ [Eq.~(\ref{eq:dg132})] to $v_2$ is weaker because of a larger spread in the cluster angular correlation $\mean{\cos(\phi_\alpha-3\phi_\beta+2\phi_\cl)}$ by the additional $2(\phi_\alpha-\phi_\beta)$. This is indeed what is observed in the ALICE~\cite{ALICE:2020siw} and STAR~\cite{STAR:2025uxv} data.

\subsubsection{Event-shape engineering}
The major flow-induced background to the CME $\dg$ measurements, Eq.~(\ref{eq:cluster}), is proportional to the elliptic flow $v_2$. 
One may therefore try to select events with varying $v_2$ while keeping the possible CME signal unchanged. 
This can be achieved through the so-called event-shape engineering (ESE)~\cite{Schukraft:2012ah} within a narrow centrality bin. 
The narrow centrality bin approximately fixes the number of spectators, which primarily
determines the magnetic field, thus the CME signal. 
The elliptic flow still varies event-by-event because of the initial geometry fluctuations.
In this approach, the event selection is based on the reduced flow vector,
\begin{equation}\label{eq:q}
    \mathbf{q}_2=\sqrt{N}\left(\mean{\cos2\phi},\mean{\sin2\phi}\right),
\end{equation}
calculated with particles (or more generally from energy flow) in a given momentum region, usually at forward or backward rapidities.
It was shown that the magnitude, $q_2$, is closely related to the average elliptic flow $\mean{v_2}$ of POIs in another momentum region, usually at midrapidity, to avoid autocorrelation between $q_2$ and flow measurements. 
The $q_2$ magnitude varies event-by-event primarily due to statistical fluctuations because of the finite number of particles or energy measurements used to compute $q_2$. 
These statistical fluctuations cancel in the mean $\mean{v_2}$ of POIs because $q_2$ and $v_2$, being from different momentum regions, are statistically independent. The variation in $\mean{v_2}$ over the $q_2$-selected event classes is dynamical, due presumably to the fluctuating initial geometry enabled by the $q_2$ selection of events.
One then studies the physics of interest (in our case the CME) as a function of the POI $\mean{v_2}$ from the $q_2$-selected event classes.  
Typically this variation in $\mean{v_2}$ is within a factor of two~\cite{Acharya:2017fau}, and the probability of events with small/large $\mean{v_2}$ extremes is low. 
This results in a relatively large statistical uncertainty in the extrapolated intercept of $\dg$ measurement as a function of $\mean{v_2}$, the primary signal of interest in the ESE method.
The results obtained with this method are presented in Sect.~\ref{sec:result:ESE}.

On the other hand, the statistical fluctuations in the apparent event-by-event anisotropy quantities, e.g.~the $v_2^\obs\equiv \mean{\cos(2\phi-2\psiep)}$~\cite{STAR:2013zgu}) and the $q_2$ of single particles as well as particle pairs~\cite{Xu:2023elq,Xu:2023wcy}, are large.
It is tempting to use these quantities calculated from the POIs themselves to select events where the background contribution is small due not only to the small(er) ellipticity of the initial geometry but also to statistical fluctuations in particle distribution in a given event. 
The so-called event-shape selection (ESS) method is based on this idea. 
In such selected events the mean $v_2$ of POIs is dominated by statistical fluctuations because the event-selection variable is based on POIs. 
In events where the overall POIs' $v_2=0$, the $v_2$ values of the background-contributing sources are not necessarily zero~\cite{Wang:2016iov}. 
Such biases are difficult to disentangle, as it would inevitably require the full knowledge of all underlying components of the event.
Further quantitative investigations are required to clarify the practical applicability of the ESS method in CME searches.

\subsubsection{Isobar and proton-nucleus collisions}
The ideal scenario to disentangle the CME signal from backgrounds is to find a pair of systems where the backgrounds are identical whereas the CME signals differ. This is the idea behind the proposal of isobar collisions~\cite{Voloshin:2010ut}.
A pair of isobar nuclei have the same number of nucleons and different numbers of protons. 
The magnetic field (and, correspondingly, the CME signal) is thus expected to be different, whereas the background due to flow is expected to be very similar. 
A dedicated isobar program using $^{96}_{44}$Ru and $^{96}_{40}$Zr was conducted at RHIC in 2018.
The central objective is to measure the double ratio between the two isobar systems, $(\dg/v_2)_{\Ru}/(\dg/v_2)_{\Zr}$.  
Note that using the ratio $\dg/v_2$ has two advantages: 1) its measurement does not require knowledge of the EP resolution~\cite{Voloshin:2018qsm}, which reduces the systematic uncertainty of the result; 2) it normalizes the $\gamma$ correlator to the $v_2$ value (to which the background is proportional) and thus can be used for a signal comparison in isobar collisions even if the $v_2$ values are slightly different in the two systems.  
The CME fraction in \Zr\ collisions can be obtained by
\begin{equation} 
    \fcme^{\sms \Zr} = \left[ \frac{\dgv_{\Ru}}{\dgv_{\Zr}} - 1 \right] \left/ 
        \left[ \left(\frac{B_{\sms \Ru}}{B_{\sms \Zr}}\right)^2-1 \right] \right.\,.
\label{eq:dblratio}
\end{equation}
$B_{\sms \Ru}/B_{\sms \Zr}$ is the ratio of the magnetic field strengths in \Ru\ and \Zr\ collisions, and can be taken as the ratio of the nuclear charges or obtained from theoretical calculation which should be rather robust despite many uncertainties on the magnetic field strengths in the individual systems.

The naive expectation is that the double ratio $(\dg/v_2)_{\Ru}/(\dg/v_2)_{\Zr}$ would be greater than unity, because any CME signal would be larger in \Ru\ collisions than in \Zr\ collisions~\cite{Voloshin:2010ut,Deng:2016knn}.
However, the experimental results~\cite{STAR:2021mii} did not align with this expectation, indicating a non-negligible background difference between the isobars, which will be discussed in detail in Sect.~\ref{sec:result:isobar}. 

A complementary strategy is to compare 
proton–nucleus (pA) and nucleus–nucleus (AA) collisions at the same final-state multiplicity, where similar collective flow has been observed~\cite{Dusling:2015gta}.
In pA collisions, the magnetic field direction is largely uncorrelated with the event geometry, effectively turning off the CME contribution while preserving comparable flow backgrounds~\cite{CMS:2016wfo,Belmont:2016oqp}.
By matching the multiplicity (or $v_2$) between pA and AA events, one can test whether the charge-dependent correlations scale with the magnetic field (as expected for CME) or with flow (as expected for background). One shortcoming of this strategy as compared to isobar collisions is that the backgrounds may still be different between pA and AA even at the same multiplicity (or $v_2$). 

\subsubsection{Comparative measurements with spectator/participant planes} \label{sec:method:ppsp}
The flow-induced background is the largest in $\dg$ measurement with respect to the second-order flow plane (i.e., PP) 
and is smaller in that with respect to the SP by a factor given by the ratio of the respective $v_2$ values. 
Because the magnetic field is determined primarily by spectator protons, the CME-induced $\dg$ would be the largest with respect to the SP and would be reduced in magnitude if measured relative to the PP, likely by the same factor of the $v_2$ ratio, namely
\begin{equation}
  \vtwosp/\vtwopp = \dg_\cme^\spp/\dg_\cme^\ssp =
  \mean{\cos2(\psi_\spp-\psi_\ssp)}\equiv a.
\end{equation}
The elliptic flow values measured with respect to the PP, $\vtwopp$, and to the SP, $\vtwosp$, 
differ by an appreciable 10-20\% depending on the collision centrality.  
This provides an opportunity to extract the CME signal fraction from the two $\dg$ measurements with respect to SP and PP, in place of $\psi_\srp$ in Eq.~(\ref{eq:gamma}), by~\cite{Xu:2017qfs,Voloshin:2018qsm}
\begin{equation}\label{eq:fcme_obs}
    \fcme^\spp\equiv\frac{\dg_\cme^\spp}{\dg^\spp}=\frac{A/a-1}{1/a^2-1}\,,
\end{equation}
where $A=\dgsp/\dgpp$.  
Note that the calculation of the double ratio, $A/a=(\dgsp/\vtwosp)/(\dgpp/\vtwopp)$, does not require knowledge of the EP resolutions and can be measured more accurately than, e.g., the ratio $A$. 
The deviation of the double ratio from unity would immediately indicate the presence of the CME contribution.

The flow-induced background is removed by design in this SP/PP
method~\cite{Xu:2017qfs,Voloshin:2018qsm}. 
The method is unique in the sense that it does not depend on the specific details of the physics background, whether it is induced by collective flow or, for example, by spin alignment of vector mesons from color field fluctuations;
as long as the physics background contributing to $\dg$ is proportional to $v_2$, it is accounted for by the method.
The method is similar in spirit to the idea of isobar collisions~\cite{Voloshin:2010ut} -- both approaches  compare two measurements that supposedly differ in $\fcme$. 
The SP/PP method has an advantage in that the collision events used for SP and PP measurements are identical, whereas the background in the isobar collisions might have subtle differences, due to differences in the centrality selections and/or isobar nuclear structures~\cite{Xu:2017zcn}.

\section{Experimental results and discussion}\label{sec:result}

\subsection{Early measurements}
\begin{figure*}[hbt]
  $\vcenter{\hbox{\includegraphics[width=0.45\textwidth]{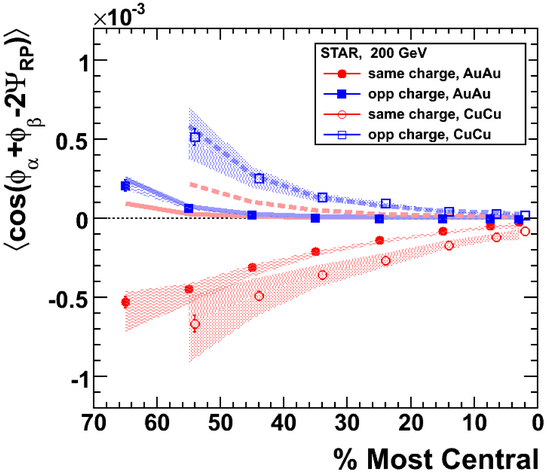}}}$\hfill
  $\vcenter{\hbox{\includegraphics[width=0.5\textwidth]{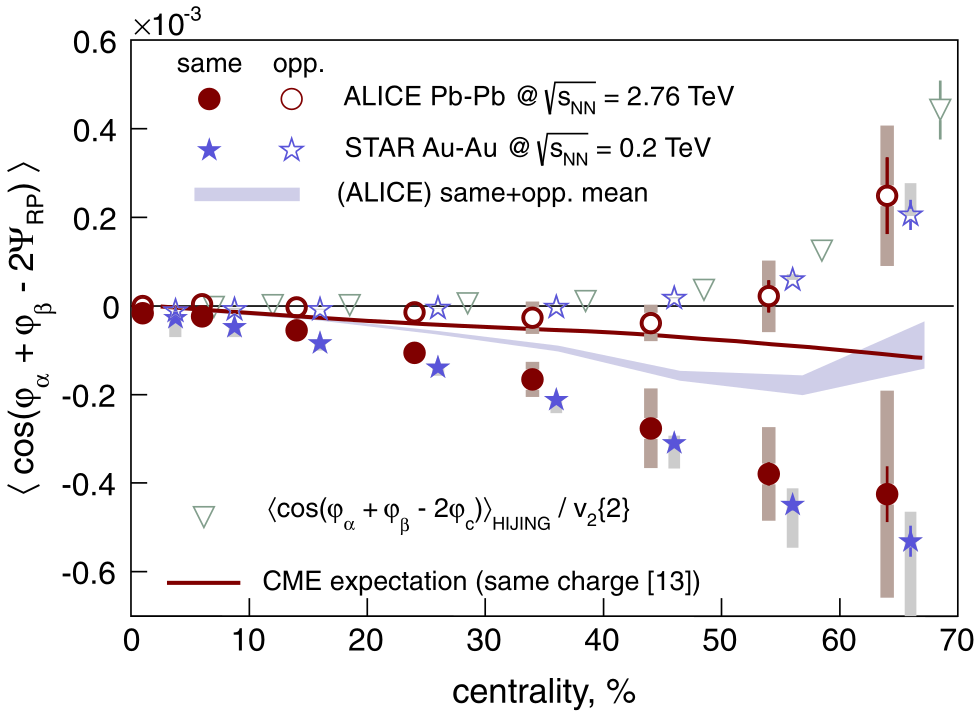}}}$
  \caption{(Color online) The first measurements
    of the opposite-sign (OS) and same-sign (SS) $\gamma$
    correlators in Au+Au and Cu+Cu collisions at
    $\snn=200$~GeV by STAR (left panel)~\cite{STAR:2009wot,STAR:2009tro} and in Pb+Pb collisions at $\snn=2.76$~TeV by ALICE (right panel)~\cite{Abelev:2012pa}. 
    See Refs.~\cite{STAR:2009wot,STAR:2009tro,Abelev:2012pa} for more details. Plots are taken from Refs.~\cite{STAR:2009wot,STAR:2009tro} and Ref.~\cite{Abelev:2012pa}, respectively.}
\label{fig:first}
\end{figure*}

The first results on the CME search were published by the STAR Collaboration in 2009~\cite{STAR:2009wot,STAR:2009tro}. As shown in Fig.~\ref{fig:first}, a clear difference is observed between the correlators $\gos$ and $\gss$ in Au+Au and Cu+Cu collisions at $\snn=200$~GeV. The $\gos$ and $\gss$  results are not symmetric about zero, as would be expected from a pure CME scenario. However, they were consistent with the general expectation of a CME signal on top of a common negative background, which could be attributed to charge-independent correlations.
STAR has also measured the $\gamma$ correlators with respect to the first-order harmonic plane $\psi_1$ from the zero-degree calorimeters (ZDCs) determined by the spectator neutrons~\cite{STAR:2013ksd}.  
The results were generally consistent with the measurements with respect to the TPC $\psi_2$ with sizeable uncertainties. 

The ALICE experiment at the LHC measured the $\gamma$ correlator in Pb+Pb collisions at $\snn=2.76$~TeV~\cite{Abelev:2012pa}.  
The results are shown in Fig.~\ref{fig:first}, and are close in magnitude to the STAR measurements at RHIC and exhibit similar trends.  
While similar trends are generally expected, the similar magnitudes are surprising, given the differences in the average multiplicities, lifetimes of the magnetic field, values of the elliptic flow, and somewhat different acceptances. 
Because the $\dg$ correlators are dominated by flow-induced background contributions, the similarity is most likely accidental.

\begin{figure}[hbt]
  \begin{minipage}{0.6\textwidth}
    \includegraphics[width=\textwidth]{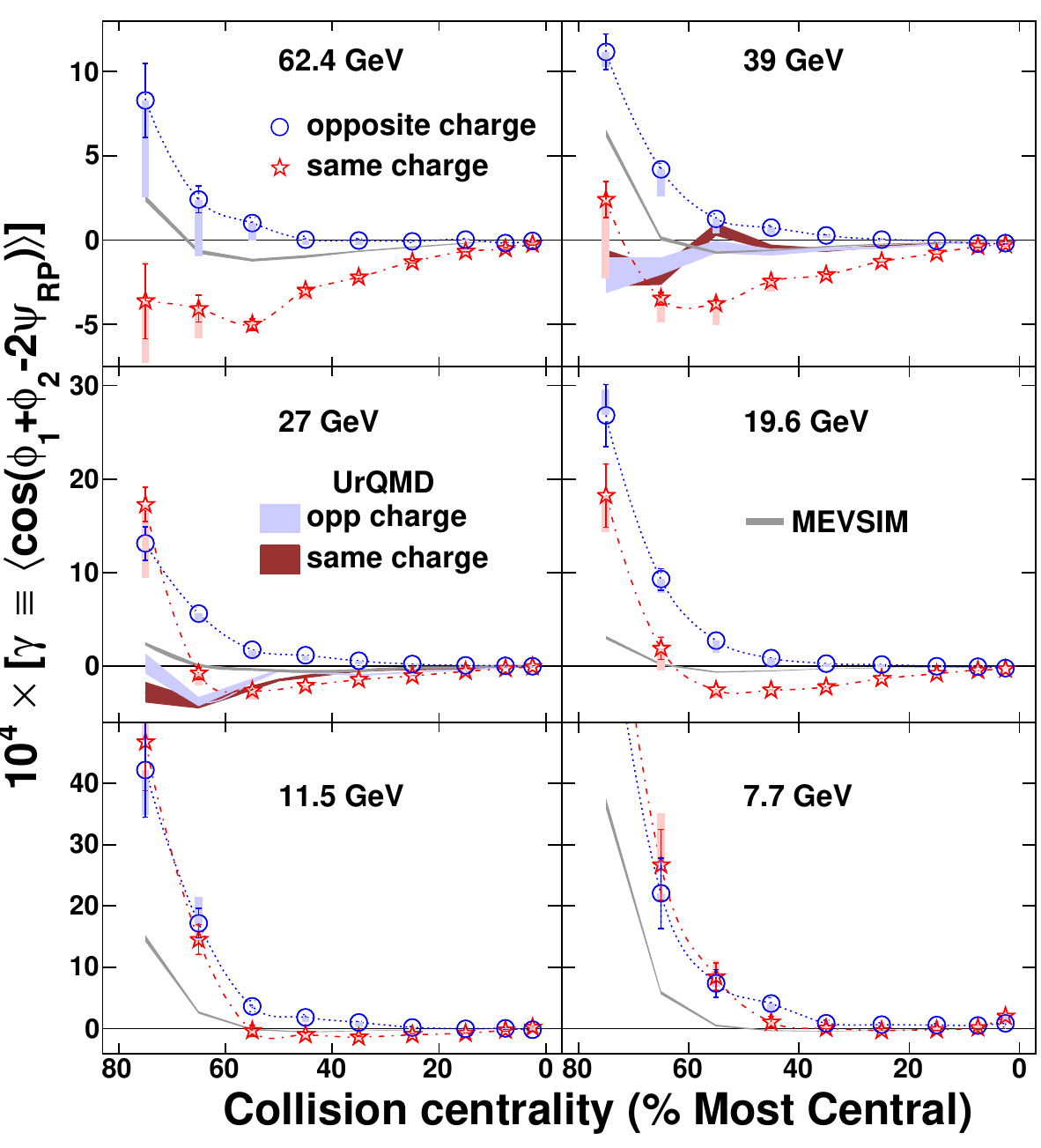}
  \end{minipage}\hfill
  \begin{minipage}{0.35\textwidth}
    \vspace*{3.5cm}
    \caption{\label{fig:STAR_BES}(Color online) 
      The three-point correlators $\gos$ and $\gss$ as a function of centrality for Au+Au collisions at $\snn=7.7$-–62.4~GeV~\cite{STAR:2014uiw} from STAR. 
      Figure is taken from Ref.~\cite{STAR:2014uiw}.}
  \end{minipage}
\end{figure}

STAR has also measured the $\gamma$ correlators at lower RHIC energies from
the Beam Energy Scan phase-I (BES-I) data~\cite{STAR:2014uiw}.  
These results are shown in Fig.~\ref{fig:STAR_BES}. 
Differences between $\gos$ and $\gss$ were observed varying with energy, disappearing towards the measured lowest energies. 
This most likely reflects quantitative changes of background contributions with energy, rather than the possible CME signal.

\subsection{Evidence of backgrounds}
\begin{figure}[hbt]
  \includegraphics[width=0.5\textwidth]{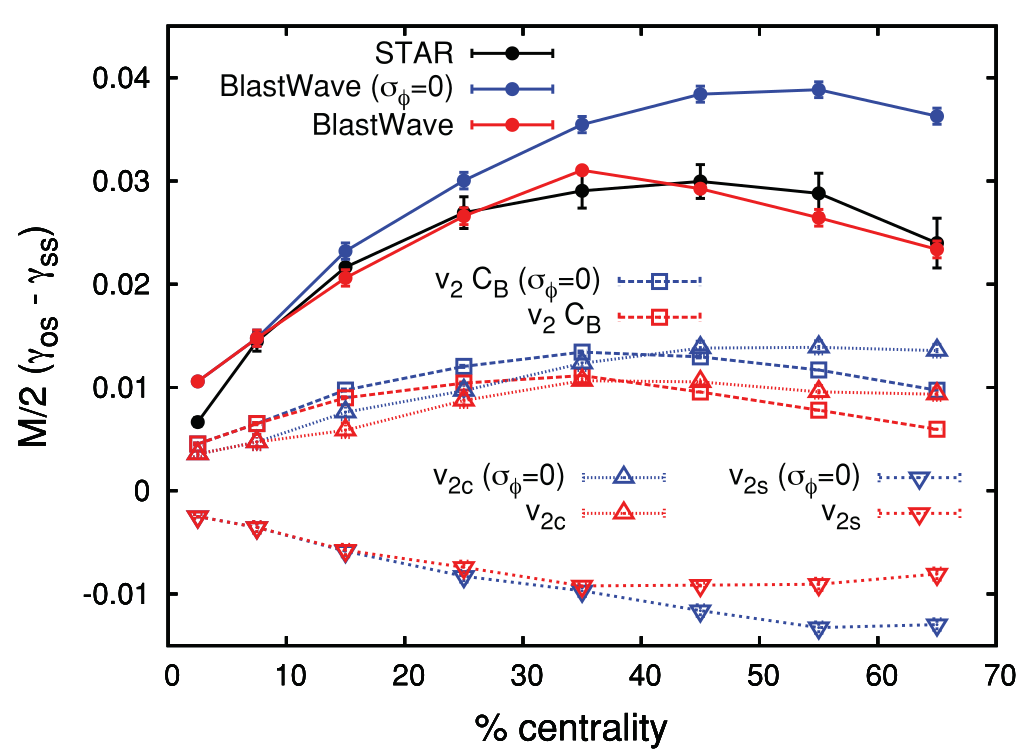}\hfill
  \includegraphics[width=0.5\textwidth]{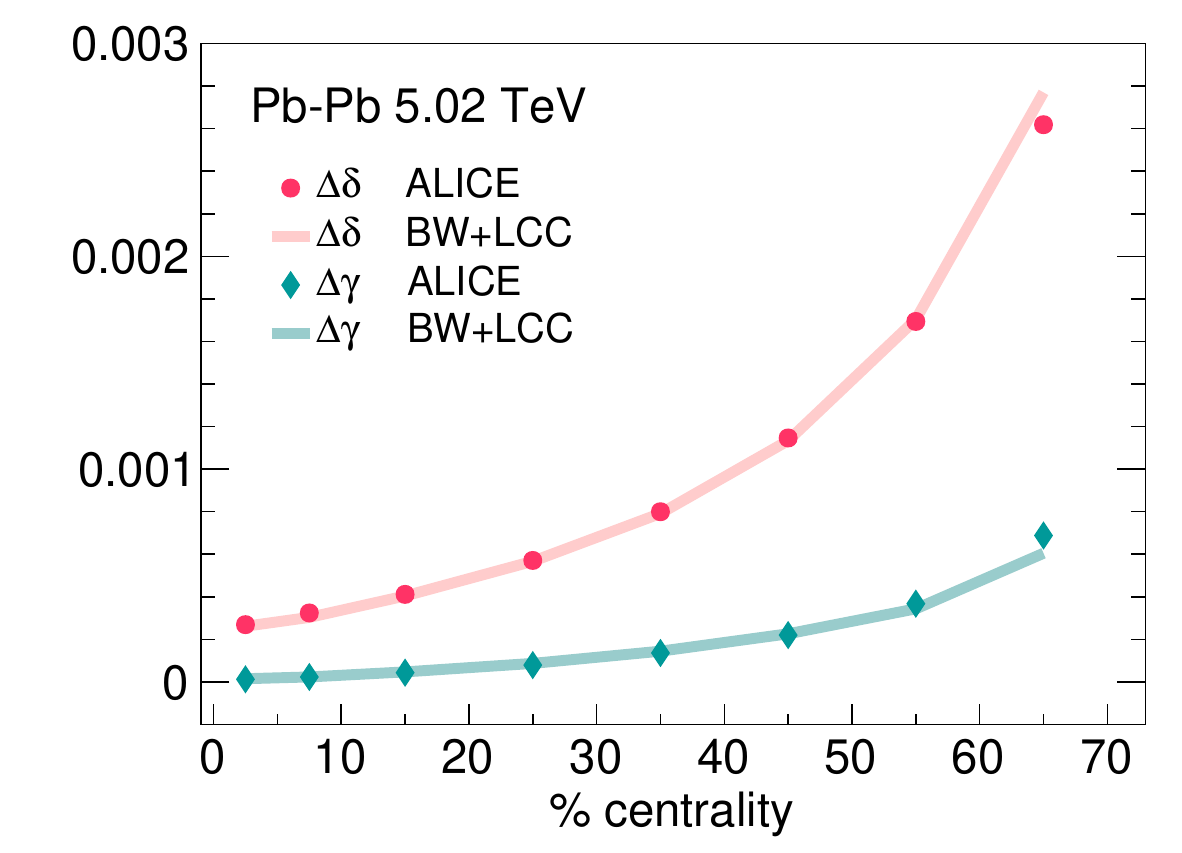}
  \caption{\label{fig:BW}(Color online) Upper panel: The multiplicity-scaled $\dg$
    from Blast-Wave model calculations~\cite{Schlichting:2010qia}
    for realistic LCC effect at freeze-out (red dots) and
    perfect local charge conservation (blue dots), compared to the STAR data (black dots). For explanation of the dashed curves, presenting separate contributions, see Ref.~\cite{Schlichting:2010qia}. Plot is taken from Ref.~\cite{Schlichting:2010qia}. Lower panel: The BW+LCC description for the ALICE data. Plot is taken from Ref.~\cite{Wu:2022fwz}.}
\end{figure}

The existence of the flow-induced background was known and thought to be significant~\cite{Voloshin:2004vk,Wang:2009kd,Bzdak:2009fc,Schlichting:2010qia}. 
The Blast-Wave (BW) calculations~\cite{Schlichting:2010qia}, including the LCC and based on the parameterization of the momentum spectra and elliptic anisotropy data of Au+Au collisions at RHIC, can reproduce the STAR measurements; see Fig.~\ref{fig:BW} top.
The LCC picture adopted in~\cite{Schlichting:2010qia} assumes that the pairs of opposite charges are created very close in space at the late stage of the system evolution. 
Radial boost of the pair due to transverse expansion leads to particle collimation in azimuth and pseudorapidity~\cite{Voloshin:2003ud,Voloshin:2004th}. 
Due to elliptic flow, the collimation is stronger in-plane than out-of-plane, which contributes to $\dg$~\cite{Schlichting:2010qia}.

The description of the BW+LCC model has also been extended to Pb+Pb collisions at the LHC~\cite{ALICE:2020siw,ALICE:2022ljz}. As shown in the lower panel of Fig.~\ref{fig:BW}, when the model is tuned to reproduce three key observables, multiplicity, $v_2$, and balance functions, it can simultaneously and accurately describe not only CME-sensitive $\dg$, but also the chiral magnetic wave (CMW) observable~\cite{Wu:2022fwz}. The CMW is another important higher-order chiral anomalous effect~\cite{Burnier:2011bf,Kharzeev:2010gd}, quantified experimentally through the correlation between the event charge asymmetry and $v_2$~\cite{STAR:2015wza,ALICE:2015cjr,CMS:2017pah,Xu:2019pgj,STAR:2022zpv,ALICE:2023weh}. This unified description further offers a phenomenological approach for integrating CME and CMW signals, enabling an joint estimate of their maximum allowable strength, which is found to be small.

It is important to note, however, that the success of these BW descriptions only suggest the major background contribution from LCC, but does not conclusively rule out the signal. This is because 
the BW parameters were {\em tuned} to the measured charge balance function, whose detailed shape could already contain a possible CME signal.

\begin{figure*}
  $\vcenter{\hbox{\includegraphics[width=0.5\textwidth]{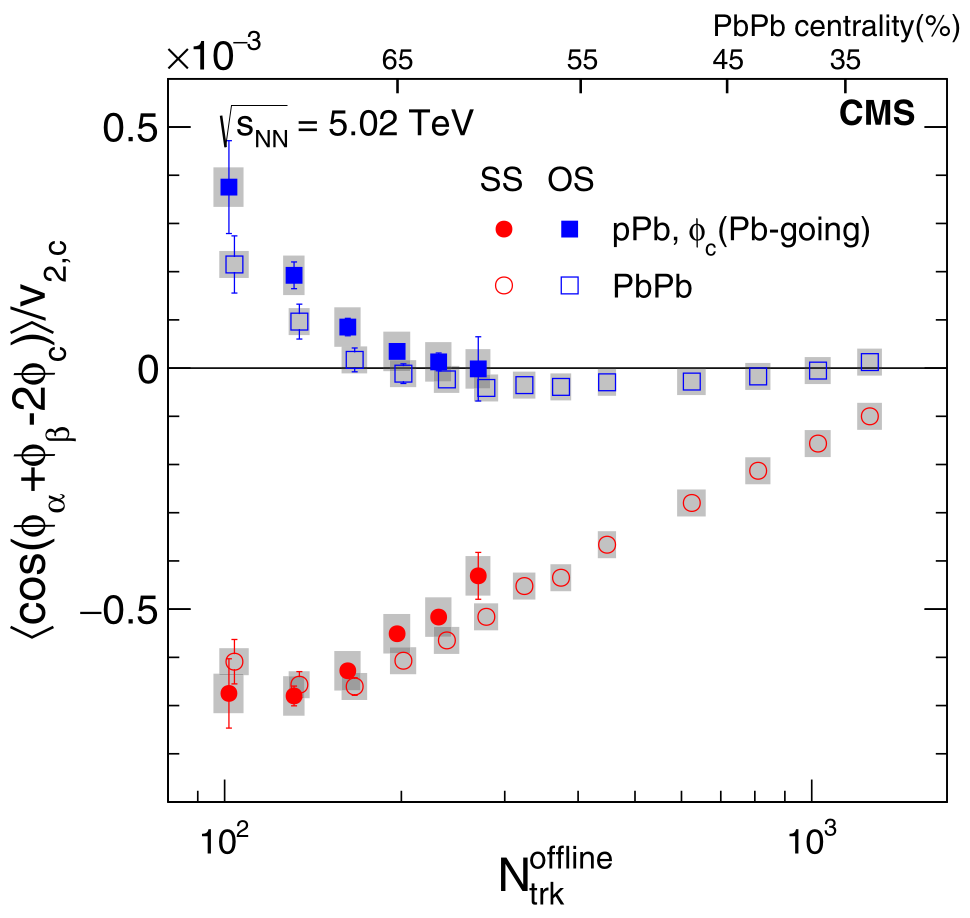}}}$\hfill
  $\vcenter{\hbox{\includegraphics[width=0.5\textwidth]{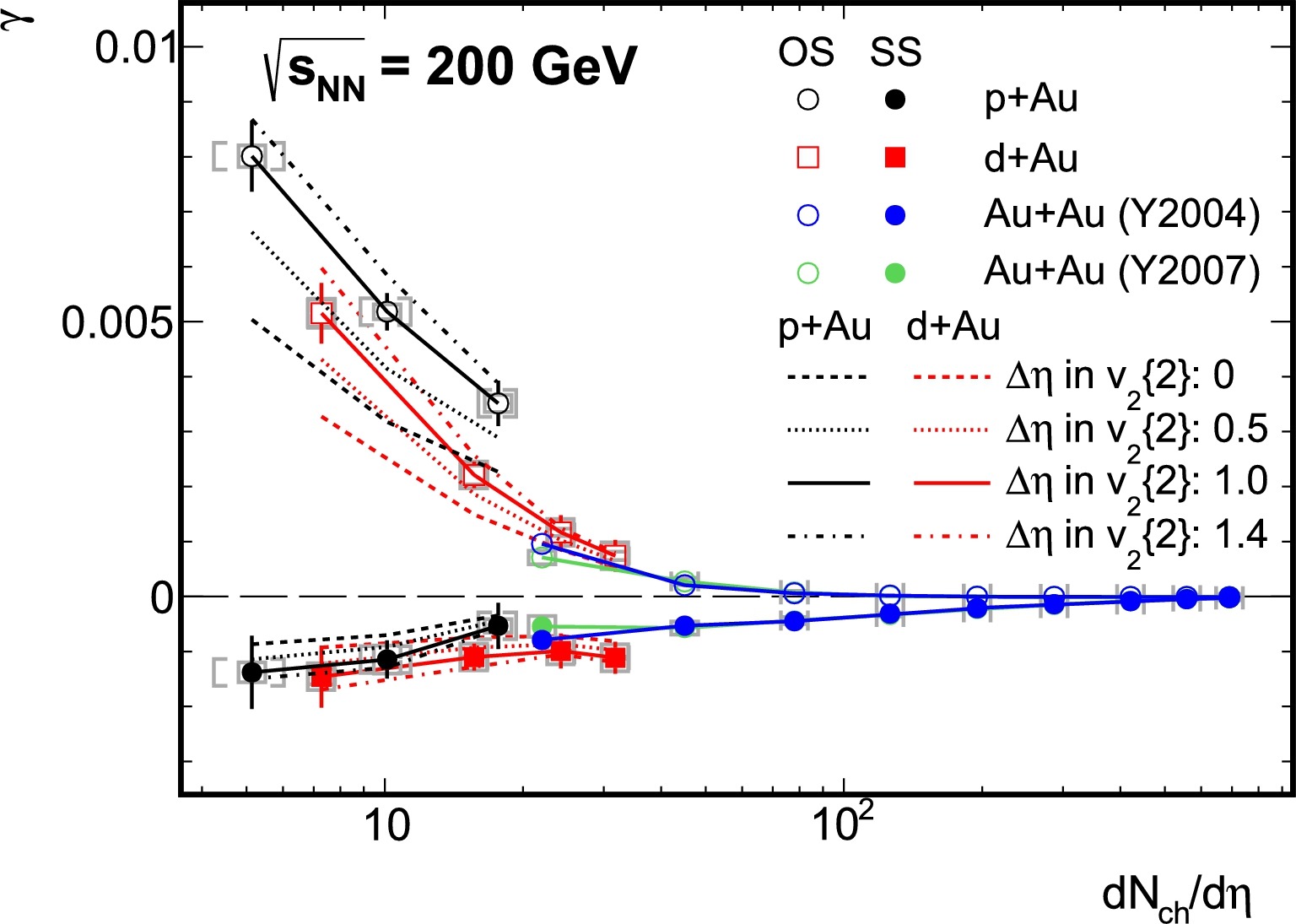}}}$
  \caption{(Color online) 
    The $\gos$ and $\gss$ correlators in small systems compared to those in large systems as functions of multiplicity in $p$+Pb and Pb+Pb collisions at $\snn=5.02$~TeV by CMS (left panel)~\cite{CMS:2016wfo}, 
   and in $p$+Au, $d$+Au, and Au+Au  collisions at $\snn=200$~GeV by STAR (right panel)~\cite{STAR:2019xzd}. 
    Plots are taken from Ref.~\cite{CMS:2016wfo} and Ref.~\cite{STAR:2019xzd}, respectively.
    }
  \label{fig:small}
\end{figure*}

The first unambiguous experimental evidence of backgrounds dominating the $\dg$ observable is from measurements in small systems, where the CME signal is expected to be negligible. 
Such measurements have been performed in p+Pb collisions by the CMS Collaboration at the LHC~\cite{CMS:2016wfo}, and in $p$+Au and $d$+Au collisions by the STAR Collaboration at RHIC~\cite{STAR:2019xzd}. 
These results are shown in Fig.~\ref{fig:small}, where large differences between $\gos$ and $\gss$ are observed, comparable to those measured in peripheral heavy ion collisions at similar multiplicities. 
Because the magnetic field direction is not correlated with the PP direction in small systems, which arises from pure fluctuation effects, any CME signals, even existent in those small-system collisions, are not observable by the $\dg$ correlator~\cite{CMS:2016wfo,Belmont:2016oqp}.
The observed $\dg$ in small systems and peripheral heavy ion measurements are likely dominated by RP-independent three-particle correlations, as suggested by \hijing~model calculations~\cite{Zhao:2019kyk}.

\begin{figure*}
  \centering
  \begin{minipage}{0.46\textwidth}
    \includegraphics[width=\textwidth]{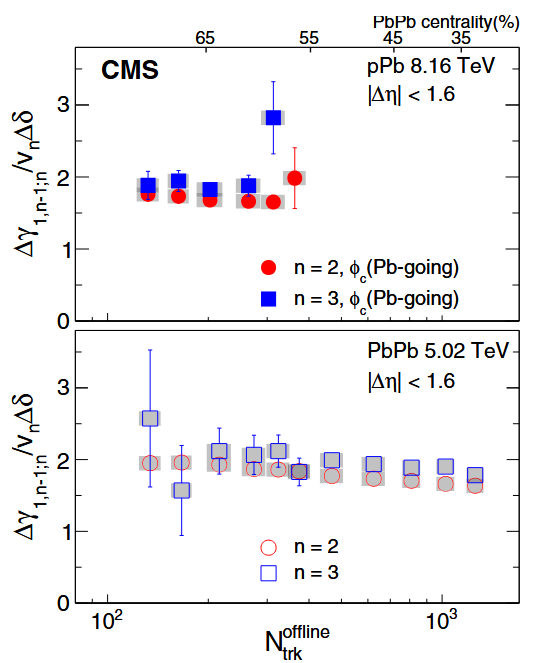}
    \caption{(Color online) The $\kappa_n$ ($n$=2,3) parameters measured by CMS in 8.16~TeV p+Pb (upper panel) and 5.02~TeV Pb+Pb (lower panel) collisions. 
      Figure is taken from Ref.~\cite{CMS:2017lrw}.}
    \label{fig:kappa}
  \end{minipage}\hspace{0.03\textwidth}
  \begin{minipage}{0.49\textwidth}
    \includegraphics[width=\textwidth]{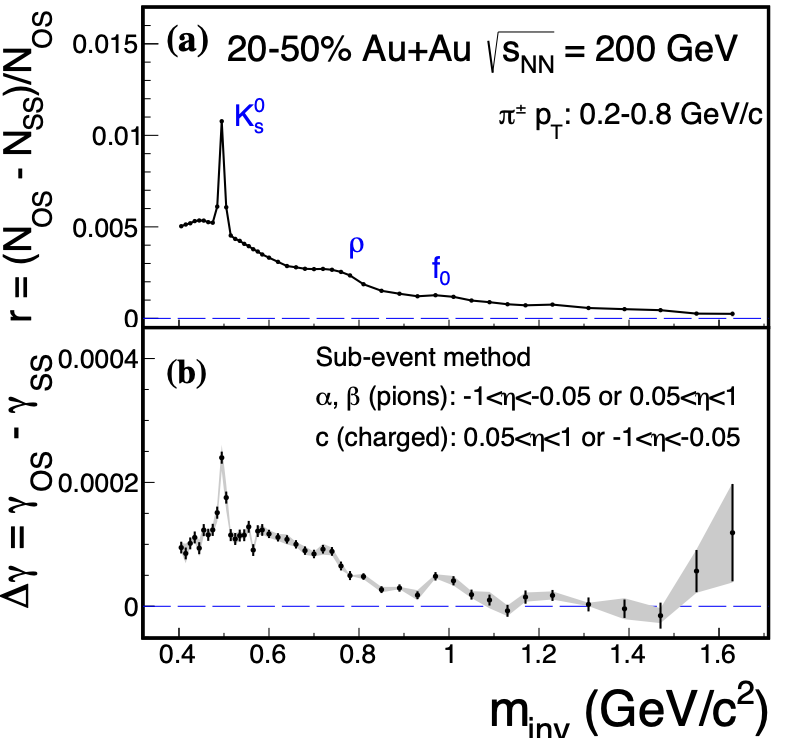}
    \caption{(Color online) The relative excess of OS over SS pion pairs (a) and the $\dg$ (b) in 20-50\% Au+Au collisions at $\snn=200$~GeV as functions of the pair invariant mass ($\minv$) measured by STAR. Figure is taken from Ref.~\cite{STAR:2020gky}.}
    \label{fig:minv}
  \end{minipage}
\end{figure*}

Figure~\ref{fig:kappa} shows the measurements of the $\kappa_n (n=2,3)$ parameters, defined by Eqs.~(\ref{eq:k2}) and~(\ref{eq:k3}), in p+Pb and Pb+Pb collisions. The $\kappa_n$ values deviate from unity, and are remarkably similar between the two collision systems. The nonzero $\kappa_3$, where no CME signal is expected, is a clear demonstration of flow-induced background dominant in the $\dg_{123}$ correlator. The CMS data further show $\kappa_2\approx\kappa_3$ within uncertainties, indicating
that both $\dg_{112}$ and $\dg_{123}$ are dominated by the same flow-driven background mechanism.
Nevertheless, as discussed in Sect.~\ref{sec:method:kappa}, while their similarity supports the background interpretation, $\kappa_3$ cannot be directly used as a precise quantitative estimator for the background component of the CME-sensitive $\dg_{112}$ correlator.

Further experimental evidence of background is observed in the $\dg$ measurement as a function of the POI pair invariant mass ($\minv$)~\cite{Zhao:2017nfq,STAR:2020gky}. This is shown in Fig.~\ref{fig:minv}, where the upper panel presents the relative difference in the OS and SS pair multiplicities and the lower panel presents the $\dg$ correlator as a function of $\minv$. Correspondences of the $K_S^0$, $\rho^0$, and $f_0$ resonance peaks between the two panels are evident; the continuum underneath the resonance peaks is suggestive of the LCC type correlations.

\subsection{Recent results}

\subsubsection{Event-shape engineering}
\label{sec:result:ESE}
\begin{figure*}[hbt]
  $\vcenter{\hbox{\includegraphics[width=0.5\textwidth]{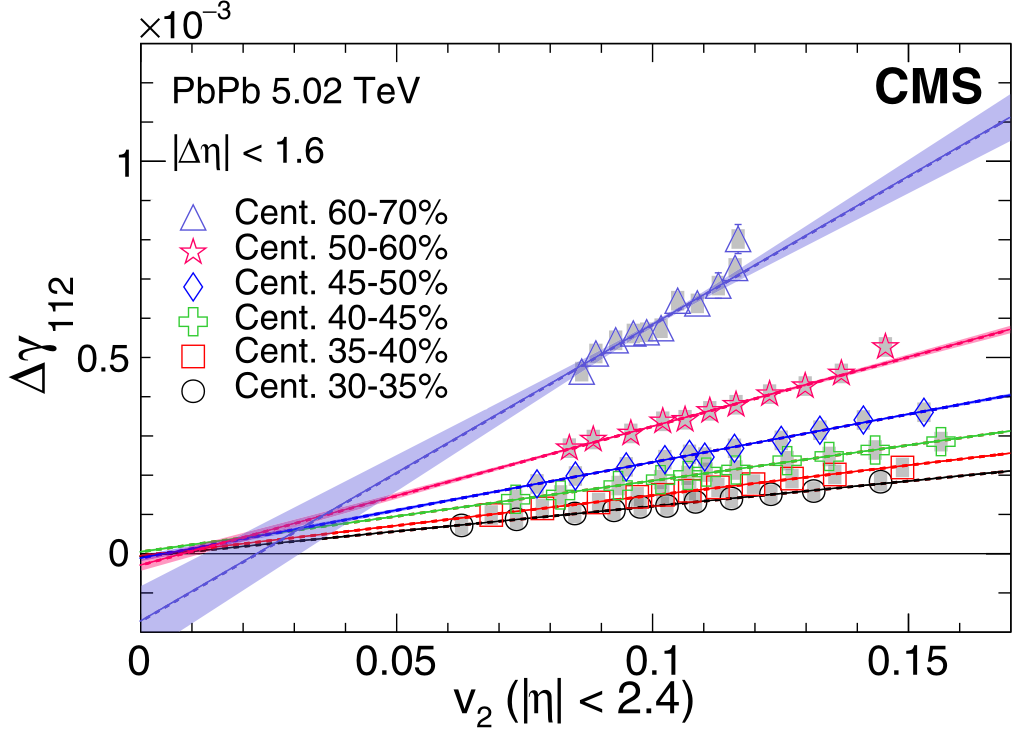}}}$ \hfill
  $\vcenter{\hbox{\includegraphics[width=0.5\textwidth]{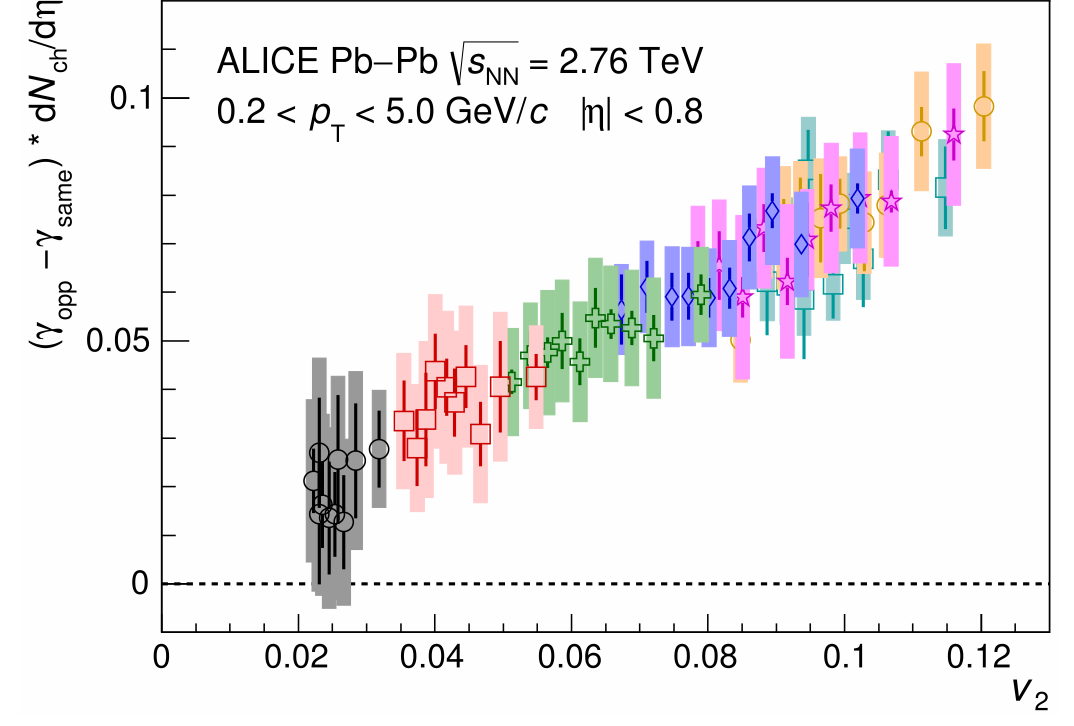}}}$
  \caption{\label{fig:ESE_LHC}(Color online) 
  Left panel: The $\dg_{112}\equiv\dg$ as a function of $v_2$ evaluated for each $q_2$ class in various centrality bins by CMS~\cite{CMS:2017lrw}. Statistical (mostly smaller than the symbol  size) and systematic uncertainties are indicated by the error bars and shaded regions, respectively. A one standard deviation  uncertainty from the fit is also show. Plot is taken from Ref.~\cite{CMS:2017lrw}. 
  Right panel: The multiplicity-scaled $d\Nch/d\eta \times \dg$ as a function of $v_2$ for ESE selected events in various centrality bins by ALICE~\cite{Acharya:2017fau}. Error bars (shaded boxes) represent the statistical (systematic) uncertainties. Plot is taken from Ref.~\cite{Acharya:2017fau}.}
  $\vcenter{\hbox{\includegraphics[width=0.5\textwidth]{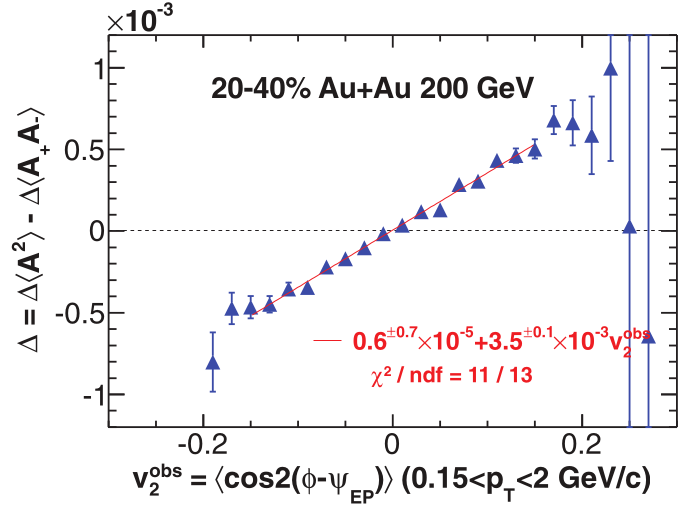}}}$ \hfill
  $\vcenter{\hbox{\includegraphics[width=0.5\textwidth]{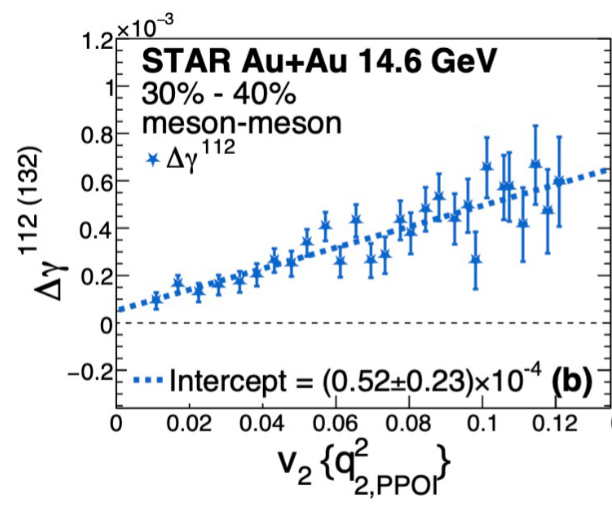}}}$
  \caption{\label{fig:STAR_ESS}(Color online) 
    Left panel: Charge multiplicity asymmetry correlations ($\Delta$, a quantity similar to $\dg$) as a function of the event-by-event $v_2^\obs$ of POIs in 20–40\% centrality Au+Au collisions at $\snn=200$~GeV by STAR~\cite{STAR:2013zgu}. Both $\Delta$ and $v_2^\obs$ are measured in the STAR TPC with respect to the first-order EP  reconstructed from the ZDCs. Plot is taken from Ref.~\cite{STAR:2013zgu}. 
    Right panel: The $\dg$ correlator as a function of single particle $v_2$ of POIs (excluding protons and antiprotons) analyzed by the ESS method in 30–40\% centrality Au+Au collisions at 14.6~GeV by STAR~\cite{Xu:2023wcy,STAR:2025uxv}.    
    The ESS method selects events based on the $q_2$ variable calculated using pairs of POIs. Error bars represent statistical uncertainties. Plot is taken from Refs.~\cite{Xu:2023wcy,STAR:2025uxv}.}
\end{figure*}

The ESE method is applied by the CMS experiment~\cite{CMS:2017lrw} in Pb+Pb collisions at $\snn=5.02$~TeV using ${q}_2$ calculated from the forward/backward hadronic calorimeters covering the pseudorapidity range of $4.4 < |\eta| < 5$, ensuring at least two units in $\eta$ away from the POIs to minimize nonflow effects.  
An approximately linear dependence on $v_2$ of the $\dg$ calculated using particles at midrapidity is observed in each centrality bin (see the left panel of Fig.~\ref{fig:ESE_LHC}). 
With the current statistical uncertainties, the CME signal (the intercept at $v_2=0$) was found to be consistent with zero~\cite{CMS:2017lrw} and an upper limit was extracted. 

The ESE method has also been applied by the ALICE experiment~\cite{Acharya:2017fau} using ${q}_2$ measured by a scintillator array detector within pseudorapidity range $-3.7<\eta<-1.7$. 
The midrapidity $v_2$ was calculated using EP reconstructed in the scintillator array detector covering $2.8<\eta<5.1$.
Figure~\ref{fig:ESE_LHC} shows the multiplicity scaled $\dg$ correlator at midrapidity as a function of $v_2$ in events binned in $q_2$ within each centrality bin. The multiplicity scaled $\dg$ removes the multiplicity dilution effect. All centralities are observed to follow a common trend, suggesting that the measured $\dg$ is dominated by flow-induced background.
Although the magnetic field and thus the CME signal is expected to be approximately constant with each narrow centrality bin, the measured $\dg$ is reduced by decorrelation of the second-order harmonic plane $\psi_2$ relative to the magnetic field direction. 
This effect was studied by ALICE using different initial geometry models and found to be dependent of the $q_2$ selection. The effect was accounted for in the ALICE analysis with a linear approximation in the measured $v_2$. 
The fit function was then used to extrapolate the $\dg$ measurements to zero $v_2$ in Fig.~\ref{fig:ESE_LHC}.  
The extrapolated intercept, sensitive to CME signal, was found to be consistent with zero~\cite{Acharya:2017fau} and  an upper limit of $\fcme<16$\% was extracted at 95\% confidence level.

More recently, the ESE method has been applied by the STAR Collaboration~\cite{Xu:2023wcy,Li:QM2023}, 
dividing the TPC acceptance ($|\eta|<1$) into three sub-events, with the midrapidity sub-event used for $q_2$ calculation and the other two, side sub-events used for POIs. The method is also applied differentially to various POI pair $\minv$ windows.
The CME-sensitive intercept extracted from this analysis is mostly consistent with zero. Because of the limited STAR TPC acceptance, the ESE analysis is challenging. Future extension would be to utilize STAR's forward capability such as the event-plane detector (EPD)~\cite{Adams:2019fpo,STAR:2022ahj}.

We note in passing that a large $\eta$ gap between POIs and the region determining $q_2$ is useful to mitigate nonflow effects, but it may reduce the power of selecting dynamical fluctuations of $v_2$ because of possible longitudinal flow decorrelations. The effects of such decorrelations are, however, expected to be small~\cite{Yan:2023ugh,STAR:2025vmb}.

It is noteworthy that all ESE analyses assumed that the effects of nonflow and RP-independent three-particle correlation backgrounds are negligible. These effects will have to be accounted for in future high-statistics measurements.

One shortcoming of the ESE method is the extrapolation to zero $v_2$, which causes large uncertainties, because of the limited range of dynamical fluctuations of $v_2$. It is attempting to exploit statistical fluctuations of $v_2$ by using the same POIs as a means to select events.
Such an attempt was first reported by the STAR Collaboration in Ref.~\cite{STAR:2013zgu}.
Figure~\ref{fig:STAR_ESS} left panel presents an example of such an analysis. 
Unfortunately, such an analysis does not fully eliminate the flow-induced background because the average $v_2$'s of background contributing sources are no longer proportional to the statistically fluctuating $v_2$ of the POIs used in the analysis~\cite{Wang:2016iov}. 
The latest attempt in this ESS approach is to use the $q_2$ of pairs of POIs for the event-shape selection.  
Example results from such an attempt is shown in the right panel of Fig.~\ref{fig:STAR_ESS}~\cite{Xu:2023wcy,STAR:2025uxv}. 
The pair $q_2$ and $v_2$, although directly containing those from the background contributing sources, are dominated by single-particle level contributions from POIs and contains a self-correlation between pairs sharing a common POI~\cite{Li:2024pue,Li:2024gdz}.
Because the POI pair $q_2$ is still selecting on the statistical fluctuations of $v_2$, 
it is challenging, if not at all impossible, to determine how much background remains in the intercept~\cite{Xu:2023elq,Li:2024gdz}. 
Experimental results using ESS are therefore not robust without further quantitative studies and modeling. 

As a reference, Table~\ref{tab:ese} contrasts the ESE and ESS methods.
\begin{table}[hbt]
\centering
\caption{Commonalities and contrasts between the ESE and ESS methods, both classifying events by flow-sensitive variables in a narrow centrality bin and projecting the $\dg(v_2)$ as a function of $v_2$ to an intercept at $v_2=0$ as a CME-sensitive measure.}
\label{tab:ese}
\begin{tabularx}{\textwidth}{m{2.6cm} *{2}{>{\centering\arraybackslash}X}}
\toprule
        &   ESE & ESS \\ \midrule
        POIs & \multicolumn{2}{c}{same} \\ \midrule
        nonflow in $\dg$ & \multicolumn{2}{c}{same, depending on $\dg$ measurement method} \\ \midrule
        event-selection variable & $q_2$ calculated not from POIs & $q_2$ calculated from POIs  (single particles or pairs) \\ \midrule
        Nonflow effects\par from event selection & depending on $\eta$ gap between POIs and $q_2$ kinematic region & likely strong effects because of no $\eta$ gap and self correlations \\ \midrule
        nature of $v_2$\par fluctuations & dynamical, from fluctuations in initial geometry and/or final-state interactions & statistical primarily, inherited from statistical fluctuations in $q_2$ \\ \midrule
        statistical\par precision\par of intercept & poor, because of small dynamical range of $v_2$ resulting in data distanced from $v_2=0$ & high, because of statistical fluctuations covering the $v_2=0$ vicinity region \\ \midrule
        premises\par and\par biases & background $v_2$ strictly proportional to final-state $v_2$, so intercept is robust as a CME-sensitive measure; biases may exist due to nonflow effects from event selection & background source $v_2$ and final-state $v_2$, both statistically fluctuating, are not guaranteed to be proportional; interpretation of intercept is unclear \\ 
\bottomrule
\end{tabularx}
\end{table}

\subsubsection{Isobar collisions}\label{sec:result:isobar}

An isobar program was conducted at RHIC in 2018 with $^{96}_{44}$Ru+$^{96}_{44}$Ru and $^{96}_{40}$Zr+$^{96}_{40}$Zr collisions at $\snn=200$~GeV~\cite{Skokov:2016yrj}. 
Exquisite control of systematics, alternating the two species on a daily basis, was maintained. 
Blind analysis~\cite{STAR:2019bjg} was performed by multiple independent research groups with different but overlapping interests and observables. 
Unprecedented precision of 0.4\% statistical uncertainty (with approximately $2\times10^9$ good MB events for each isobar species) was achieved with a negligible systematic uncertainty on the ratio of $\dg/v_2$ between the two isobar systems.
The results of the analysis were published in Ref.~\cite{STAR:2021mii}.

\begin{figure*}[hbt]
  \includegraphics[width=\textwidth]{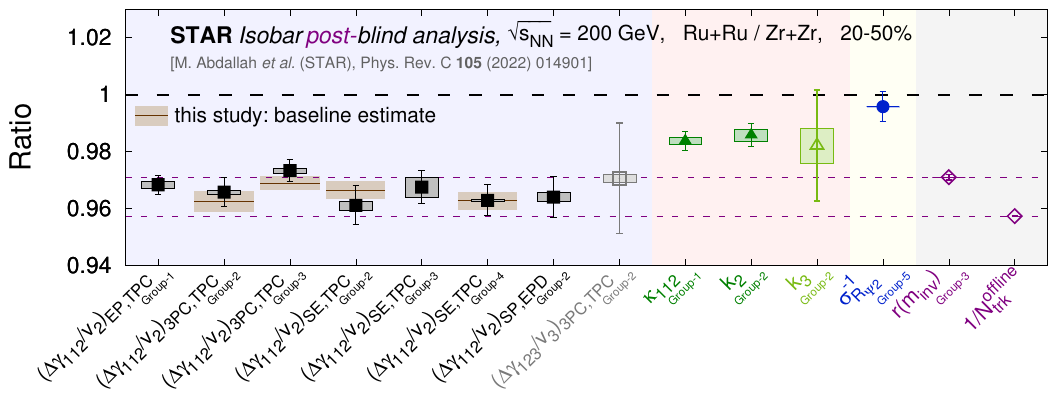}
  \caption{\label{fig:isobar}(Color online) 
  Isobar \Ru/\Zr\ ratios of $\dg/v_2$ from the STAR blind analyses (black squares, with error bars and gray boxes indicating statistical and systematic uncertainties, respectively)~\cite{STAR:2021mii}. Also shown are background baseline estimates by STAR for the four measurements that used the cumulant method (short horizontal lines accompanied by shaded boxes, the heights of which indicate the combined statistical and systematic  uncertainties)~\cite{STAR:2023gzg,STAR:2023ioo}. The rightmost and second rightmost purple diamonds indicate the isobar ratios of inverse multiplicities ($1/N_{\rm trk}^{\rm offline}$) and the relative pair excess of OS over SS pairs ($r$). 
  Figure is taken from Refs.~\cite{STAR:2023gzg,STAR:2023ioo}.}
\end{figure*}

The \Ru\ to \Zr\ ratio of the $\dg/v_2$ variable was observed to be smaller than unity (see~Fig.~\ref{fig:isobar}), contrary to the expectation of a larger CME signal in \Ru\ and the same background contributions.
The smaller-than-unity ratio results from a few percent difference in the multiplicities, greater (and thus smaller background contribution, see Eq.~(\ref{eq:cluster})] in Ru+Ru than Zr+Zr collisions. 
Such a difference in multiplicities was predicted by energy density functional theory calculations to root in the difference in nuclear structure between the $^{96}_{44}$Ru and $^{96}_{40}$Zr nuclei~\cite{Xu:2017zcn,Li:2018oec,Li:2019kkh,Xu:2021vpn}: 
the larger number of neutrons in $^{96}_{40}$Zr yields a thicker neutron skin, larger overall size, smaller energy density, and hence lower multiplicity in \Zr\ collisions. 
We note that uncertainties in determining nuclear structure from multiplicity or other measurements (mostly in low-energy nuclear reactions) are significant because of large model dependencies. 
However, the CME background is directly affected by the multiplicity, which is well measured. 
Inaccuracies in nuclear structure parameters, therefore, do not have a direct bearing on the accuracy of the relative background contributions between the two isobar systems.

As the number of clusters contributing to the background is not necessarily proportional to the final-state particle multiplicity with a percent accuracy, the uncertainty on the baseline is on the same order of the difference between inverse multiplicity scaling (the lower dashed purple line in Fig.~\ref{fig:isobar}) and the relative abundances of clusters, $r\equiv(N_\sos-N_\sss)/N_\sss$ (the upper dashed purple line in Fig.~\ref{fig:isobar}).
As can be seen, the $\dg/v_2$ isobar ratios are mostly in-between these two baseline estimates.  
Clearly, at this level of accuracy, one cannot conclude on the existence of the CME or the lack thereof, and a more accurate estimate of the background accounting for the nonflow effects must be performed~\cite{Feng:2021pgf,Feng:2024eos}.

Such an estimate of the ``baseline'' in the isobar comparison has been made by the STAR Collaboration in Refs.~\cite{STAR:2023gzg,STAR:2023ioo}. 
A 2D fitting method of two-particle correlations in $(\deta,\dphi)$ was used to estimate the (difference) in nonflow contamination in $v_2\two$ measurements.  
The three-particle nonflow contamination was estimated by the \hijing\ model following Ref.~\cite{Feng:2021pgf}. 
Since \hijing\ does not include collective flow, its entire three-particle correlations can be used as an estimate of the RP-independent three-particle correlation background~\cite{Wang:1991xy,Wang:1998bha}.  
It was found that the nonflow contribution to $v_2$ and the three-particle correlation background were both smaller in Ru+Ru collisions, presumably due to the larger multiplicity dilution in Ru+Ru collisions. 
The two effects partially cancel each other, and the net effect turns out to be slightly negative compared to the baseline estimated from $r$.
These are shown by the brown lines accompanied by shaded bars in Fig.~\ref{fig:isobar} for four of the $\dg/v_2$ isobar ratio measurements, where the cumulant method is used in measuring $v_2\two$.  
The measurements of the $\dg/v_2$ isobar ratios are consistent with the estimated background baselines, within approximately $1\sigma$ uncertainty.  
Assuming the magnetic field square difference of 15\% between the isobar systems, an upper limit of approximately 10\% for $\fcme$ is extracted for each of the four measurements at the confidence level of 95\%. 
Nevertheless, we note that the RP-independent three-particle correlations and their systematic uncertainties are estimated by models.
For reference, the highest precision data point of the $\dg/v_2$ isobar ratios in Fig.~\ref{fig:isobar} has a statistical uncertainty of 0.4\%, which corresponds to an uncertainty of approximately $3\%$ in terms of $\fcme$.

The relative signal strength in smaller nuclei collisions is likely to be significantly smaller. 
The effective magnetic field scales approximately as $Z/R^2 \sim Z/A^{2/3} \sim A^{1/3}$ (where $A$ is the mass number of the colliding nuclei), so the $\dg\propto B^2$ signal would be a factor $\sim 1.5$ smaller in isobar than Au+Au collisions. 
Accounting for the larger decorrelation of the magnetic field direction with the flow planes points to even a larger decrease~\cite{ALICE:2022ljz}. 
The lifetime of the magnetic field will also be shorter in isobar collisions. 
The background is approximately inversely proportional to multiplicity, so the background is likely a factor of two larger in isobar than Au+Au collisions~\cite{Feng:2021oub}.
Thus, the $\fcme$ fraction in smaller nuclei collisions can be  smaller by at least a factor of several.
Similar conclusions hold true for Xe--Xe collisions at the LHC~\cite{ALICE:2022ljz}. The observed $\gamma$ correlator in Xe-Xe remains comparable to that in Pb-Pb collisions, rather than being smaller as one would expect if it were mainly driven by the signal.
A reduction in the CME signal due to the final-state interactions, on the other hand, would likely be smaller in isobar collisions than in Au+Au collisions~\cite{Ma:2011uma,Deng:2018dut}. 
This would be a competing effect favoring isobar collisions; however, the magnitude of this effect is largely unknown.
Considering all these effects, the small $\fcme$ observed in  isobar collisions does not exclude a significant signal in collisions of larger nuclei.

\subsubsection{Spectator/participant planes} \label{sec:result:ppsp}
\begin{figure}[hbt]
  \begin{minipage}{0.6\textwidth}
    \includegraphics[width=\textwidth]{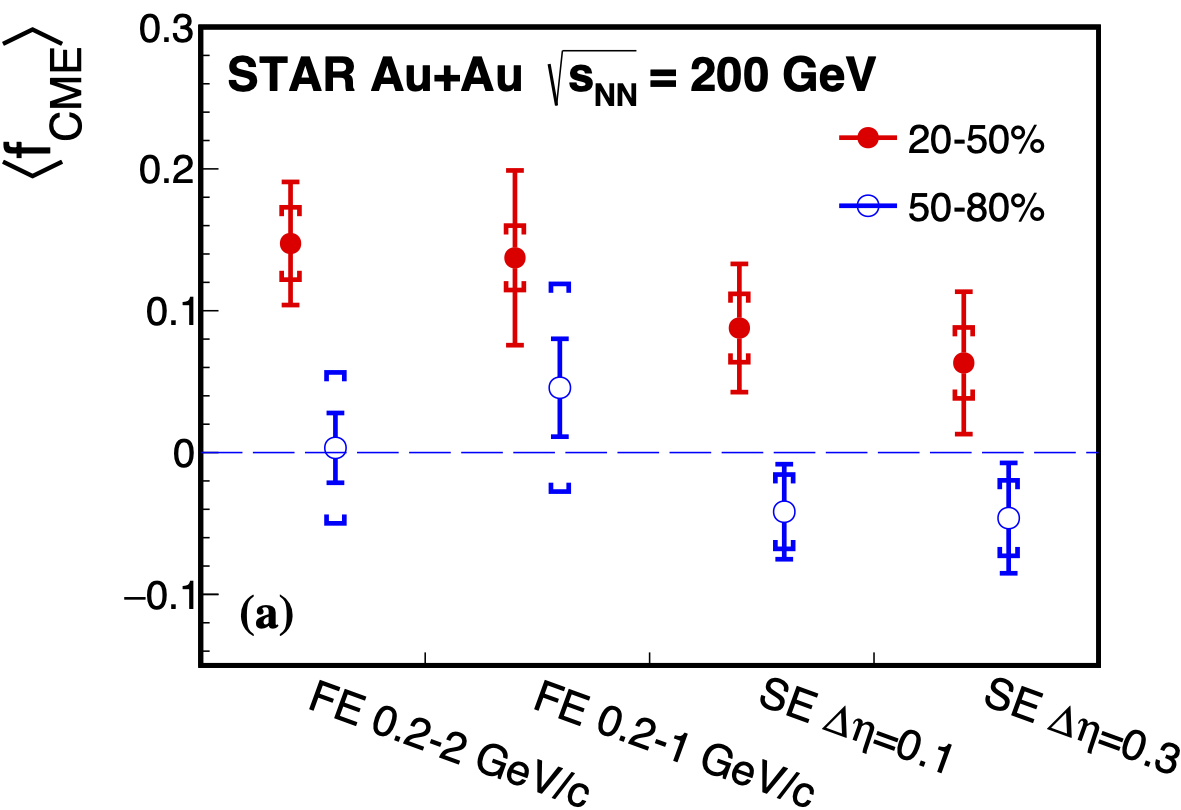}
  \end{minipage}\hfill
  \begin{minipage}{0.38\textwidth}
    \caption{\label{fig:STAR_PPRP}(Color online) 
      The observed $\mean{\fcme^\obs}$ 
      in 50–80\% (open blue markers) and 20–50\% (solid red markers) centrality Au+Au collisions at $\snn=200$~GeV by STAR~\cite{STAR:2021pwb}, extracted using methods of full-event (FE) and sub-event (SE) settings and varied kinematic cuts. 
      Error bars and caps show statistical and systematic uncertainties, respectively. 
      Figure is taken from Ref.~\cite{STAR:2021pwb}.}
  \end{minipage}
\end{figure}

The SP/PP method, described in Sect.~\ref{sec:method:ppsp}, was applied by the STAR Collaboration to Au+Au collisions at 200~GeV~\cite{STAR:2021pwb}. (Similar method was also applied to Au+Au collisions at $\snn=27$~GeV~\cite{STAR:2022ahj}.)  
The SP was estimated by the spectator neutrons measured in the ZDCs~\cite{Adams:2005ca}, and the PP by the second-order harmonic plane reconstructed from particles in the Time Projection Chamber (TPC).  
Figure~\ref{fig:STAR_PPRP} presents the measured CME fraction $\fcme^\obs$ in peripheral and midcentral collisions,  
where four data points of each correspond to four different analysis settings. 
It is found that, while consistent with zero in peripheral 50-80\% collisions, the $\fcme^\obs$ in mid-central 20-50\% collisions seems finite. 

As mentioned in Ref.~\cite{STAR:2021pwb}, the $\fcme^\obs$ measurements are still contaminated by nonflow correlations; see Sect.~\ref{sec:method:nonflow}. 
These nonflow correlations include two-particle nonflow in the $v_2\two$ measurement and RP-independent three-particle correlations in the $C_3$ measurement.
Both affect only measurements relative to the PP inferred from TPC particles, not those relative to the SP measured by the ZDCs.
The nonflow contribution to $v_2\two$ reduces $\dg\tpc$ resulting in an overestimate of the ratio $\dgsp/\dgpp$, thus a positive contribution to the $\fcme^\obs$ estimate.
The RP-independent three-particle contamination increases $\dg\tpc$ making its difference from $\dg\zdc$ smaller, thus gives a negative contribution to the $\fcme^\obs$ estimate. 
These two effects partially cancel each other. 
It is worthwhile to note that, while the experimental measurement of $\fcme^\obs$ is robust, its interpretation relies on the effect of these nonflow contamination, which is particularly important in this case because of the unconstrained sign of the effect.

\begin{figure*}
  \includegraphics[width=0.5\linewidth]{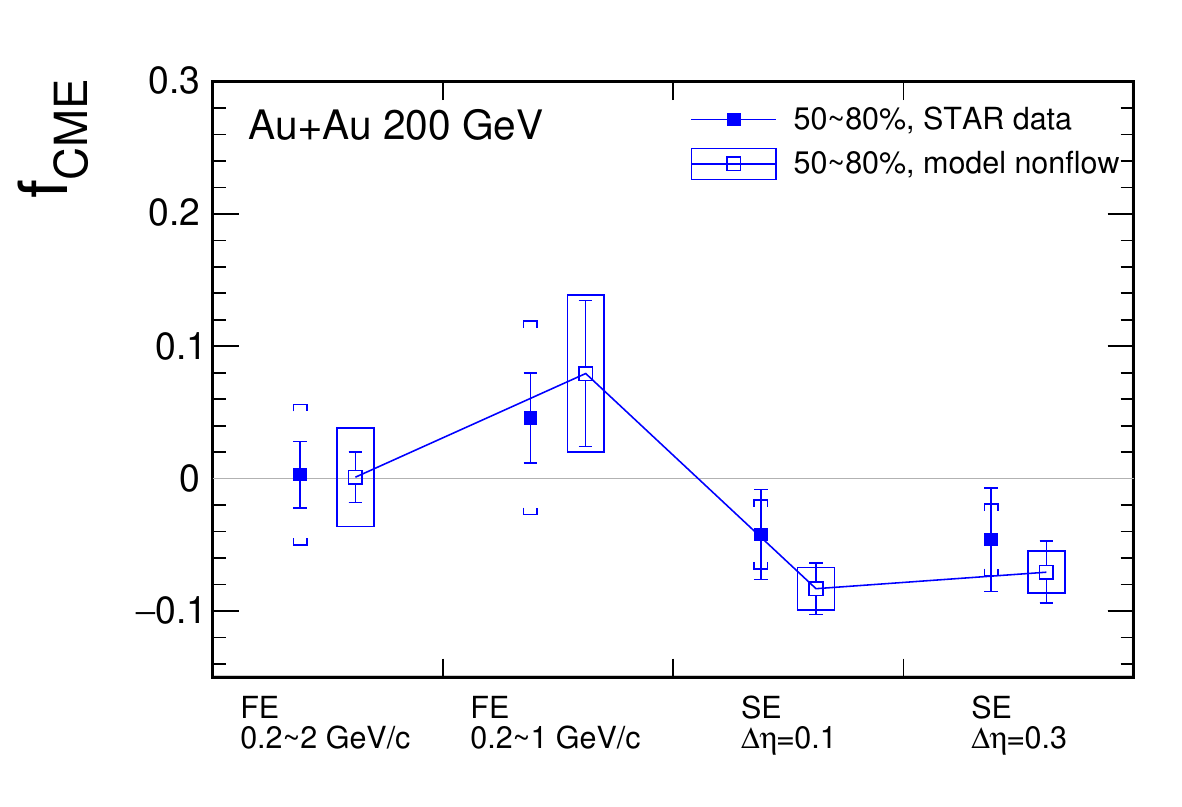}
  \includegraphics[width=0.5\linewidth]{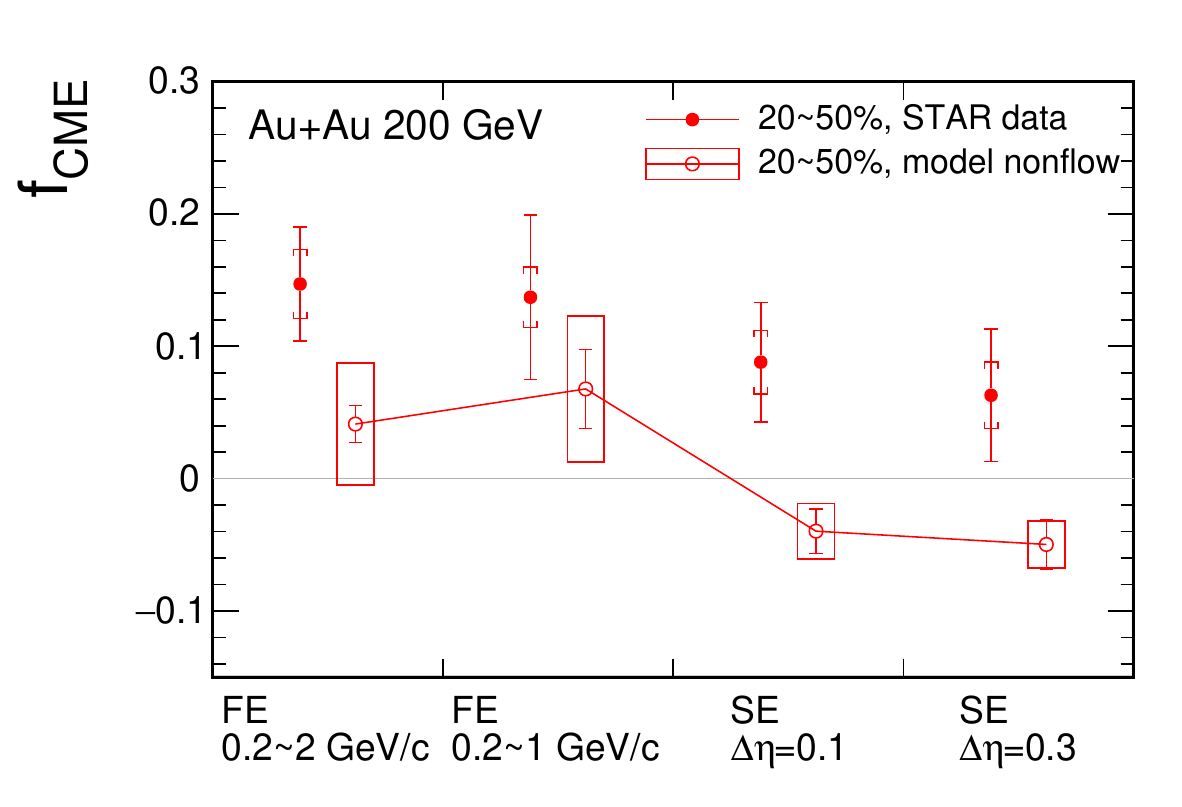}
  \caption{\label{fig:STAR_fcme}(Color online) Model estimates of nonflow  contributions to $\mean{\fcme^\obs}$ by Ref.~\cite{Feng:2021pgf}, compared to STAR data~\cite{STAR:2021pwb} for 50--80\% (left panel) and 20--50\% (right panel) centrality Au+Au collisions. 
  Error bars show statistical  uncertainties and boxes indicate systematic uncertainties on the model estimates. The STAR data are as same as in  Fig.~\ref{fig:STAR_PPRP}.
  Figure is reproduced from Ref.~\cite{Feng:2021pgf}.}
\end{figure*}

The nonflow and the RP-independent three-particle correlation effects in SP/PP measurements have been investigated in Ref.~\cite{Feng:2021pgf} using models. 
The nonflow effect in $v_2$ was estimated by using the \ampt\ model.  
The effect of RP-independent three-particle correlations is estimated by using the \hijing\ model. 
The estimated nonflow contribution to $\fcme^\obs$ is shown in Fig.~\ref{fig:STAR_fcme} by the points connected by lines, compared to the STAR measurement.  
It is found that the net effect of nonflow correlations causes a nearly zero or perhaps negative bias to $\fcme$.  
It will be interesting to have a more data-driven analysis of nonflow effects in the future.

Barring uncertainties in nonflow contamination, the current $\fcme^\obs$ observed in Au+Au collisions at $\snn=200$~GeV~\cite{STAR:2021pwb} is consistent with the isobar results~\cite{STAR:2021mii,STAR:2023gzg,STAR:2023ioo} given the expected an order of magnitude difference in the signal-to-noise ratio between Au+Au and isobar collisions as described in Sect.~\ref{sec:result:isobar}. 
It is also consistent with preliminary ESE results in 200~GeV Au+Au collisions at RHIC~\cite{Li:QM2023} and with Pb+Pb collisions at the LHC~\cite{Acharya:2017fau,CMS:2017lrw}, both consistent with zero with relatively large uncertainties as discussed in Sect.~\ref{sec:result:ESE}.
Both the SP/PP and ESE methods are affected by nonflow, in distinctive ways, so more quantitative comparisons need to await nonflow corrections. 
Nevertheless, there can be non-CME physics mimicking charge separation along the magnetic field, such as a sizable spin alignment of vector mesons or spin correlations of hyperons with respect to the global angular momentum direciton. Such non-CME physics cannot be removed by the SP/PP method, but are an integral part of the extracted signal, as they are practically identical to the CME charge-separation signature. 
\section{Summary and outlook}\label{sec:summary}
The chiral magnetic effect (CME) is a fundamental phenomenon in quantum chromodynamics arising from vacuum fluctuations violating the parity and charge-parity symmetries.  
It is expected to imprint in an electric current along the strong magnetic field in relativistic heavy ion collisions. 
The charge-dependent and reaction plane (RP)-dependent azimuthal correlator ($\dg$)~\cite{Voloshin:2004vk} is widely used to search for the CME at the Relativistic heavy ion Collider (RHIC) and the Large Hadron Collider (LHC). 

Despite extensive efforts, no conclusive evidence for the CME has been established to date. The $\dg$ correlator is found to be dominated by large background effects, arising from charge-dependent particle correlations modulated by the elliptic flow in these collisions. 
Additional backgrounds arising from nonflow correlations not related to the RP will also have to be accounted for~\cite{Feng:2021pgf}. Quantitative isolation of a possible CME signal therefore requires understanding and constraining these backgrounds at the percent level.

The RHIC isobar program of $^{96}_{44}$Ru+$^{96}_{44}$Ru and $^{96}_{40}$Zr+$^{96}_{40}$Zr collisions, designed to address the background issue~\cite{Voloshin:2010ut}, did not lead to a firm conclusion about the CME~\cite{STAR:2021mii}.  
The difficulty arises from the reduced signal-to-background ratio in relatively small systems compared to Au+Au or Pb+Pb and subtle differences in the nuclear structure of the $^{96}_{44}$Ru and $^{96}_{40}$Zr nuclei~\cite{Xu:2017zcn}.
An upper limit of approximately 10\% CME signal in the measured $\dg$ in isobar collisions is extracted at 95\% confidence level~\cite{STAR:2023gzg,STAR:2023ioo}.  
It should be pointed out that the isobar result is not a disproof of the CME but provides  guideline for its possible magnitude.  

On the other hand, an intriguing hint of a finite CME signal has been observed in Au+Au collisions at $\snn=200$~GeV at RHIC~\cite{STAR:2021pwb}. 
While nonflow effects need to be carefully investigated, this is in line with the expectation that the CME signal-to-background ratio is at least a factor of three larger in Au+Au than in isobar collisions. 
The searches for the CME at the LHC have thus far resulted only in stringent upper limits, where the CME signal may be weaker because of the faster decrease of the magnetic field at the higher energies. Table~\ref{tab:cme} juxtaposes the various constraints on the CME fraction from measurements at RHIC and the LHC.
\begin{table}
    \caption{Summary of the observed CME fraction, $\fcme^\obs$, or its upper limit at the 95\% confidence level. Except the isobar Ru+Ru results, all others have not considered nonflow contamination.}
    \label{tab:cme}
    \begin{tabular}{llrlr}
        \hline
        exp.    & system & centrality & method & $\fcme^\obs$ \\
        \hline
        ALICE   & Pb+Pb 2.76~TeV~\cite{Acharya:2017fau} & 10-50\% & ESE$^{(1)}$ & $<$ (26--33)\% \\ 
                & Pb+Pb 5.02~TeV~\cite{ALICE:2022ljz} & 0-70\% & comparative$^{(2)}$ & $(15 \pm 6)\%$ \\ 
                & Xe+Xe 5.44~TeV~\cite{ALICE:2022ljz} & 0-70\% & comparative$^{(2)}$ & $<$ 2\% \\
        \\
        CMS     
                & Pb+Pb 5.02~TeV~\cite{CMS:2017lrw} & 70-35\%    & ESE & $<$ 7\% \\
        \\
        STAR    & Ru+Ru 200~GeV~\cite{STAR:2021mii,STAR:2023gzg,STAR:2023ioo} & 20-50\% & isobar$^{(3)}$ & $<$ (8.3--11.5)\% \\
                & Au+Au 200~GeV~\cite{STAR:2021pwb} & 20-50\% & SP/PP$^{(4)}$ & $(14.7 \pm 4.3 \pm 2.6)\%$ \\
                & Au+Au 200~GeV~\cite{STAR:2021pwb} & 20-50\% & SP/PP$^{(5)}$ & $(8.8 \pm 4.5 \pm 2.4)\%$ \\
                \hline
    \end{tabular}
    \begin{itemize}
        \item[$^{(1)}$] {\small Have considered variation in $\mean{B^2\cos2(\psi_2-\psirp)}$, the effect of CME signal in the measured $\dg$, as a function of $v_2$ within each $q_2$ event class, using initial-state models.}
        \item[$^{(2)}$] {\small Compare two systems assuming CME signals to be proportional to $\mean{B^2\cos2(\psi_2-\psirp)}$, calculated with models, and backgrounds to be inversely proportional to multiplicity.}
        \item[$^{(3)}$] {\small Assume 15\% difference in $B^2$ between the two isobar systems. The quoted range for the upper limit covers four different analyses that have used the cumulant method.}
        \item[$^{(4)}$] {\small From the full-event analysis where all hadrons within $|\eta|<1$ and $0.2<\pt<2$~\gevc\ are used for the POIs ($\alpha, \beta$) and the reference particle $c$.}
        \item[$^{(5)}$] {\small From the sub-event analysis where the POIs are from $-1<\eta<-0.05$ and the reference particle $c$ is from $0.05<\eta<1$, and vice versa; the $\pt$ ranges are all $0.2<\pt<2$~\gevc.}
    \end{itemize}
\end{table}

Besides the top energy RHIC data, STAR has accumulated significant statistics at lower energies through the beam energy scan phase-I and II, primarily to search for the critical point (see, e.g., Ref.~\cite{STAR:2025zdq}). These data should be examined by the ESE and SP/PP methods. However, the statistics may still be insufficient for a significant CME search (see, e.g., Ref.~\cite{STAR:2022ahj}), especially considering the degraded ZDC performance at lower energies. 

Future CME searches must focus on achieving unambiguous background control and percent-level precision in the observable.
Promising methods to achieve this goal are the event-shape engineering (ESE) with an event-shape variable~\cite{Schukraft:2012ah} and particles of interest well distanced in momentum space and the spectator/participant planes (SP/PP) method~\cite{Voloshin:2018qsm,Xu:2017qfs}.  
The former has been successfully implemented by LHC experiments~\cite{CMS:2017lrw,Acharya:2017fau} with wide acceptance detectors and should become viable with new and future RHIC data with extended forward detector capabilities.  
The latter method has been proven to be effective at RHIC~\cite{STAR:2021pwb} owing to the good granularity of the shower maximum detectors of the zero-degree calorimeters (ZDC) and might see future applications at the LHC with improved ZDC detectors. 

STAR has recently accumulated more Au+Au collisions data at the top RHIC energy approximately ten-fold of the present statistics. 
Likewise, the increase in statistics from Run 2 to Run 3 for CME analysis in ALICE at the LHC is more than a factor of 20. 
The statistical precision will be significantly improved. 
The systematic uncertainties will likely decrease because of the increased statistics, allowing better understanding of systematic sources. 
The improved forward detector capabilities in STAR will further constrain systematic uncertainties. 
Moreover, with the increased statistics and improved detector capabilities, CME-sensitive charge separation can be studied in great detail, e.g., in different kinematic regions of $\pt$ and $\eta$.

An unfortunate choice of the RHIC isobar collision program was  the two relatively light isobar nuclei where any CME signals are likely too small to detect. 
While RHIC is ending, plenty opportunities exist at the LHC for future programs of isobar collisions involving heavier nuclei. 
Going from RHIC to the LHC, peak magnetic field, vacuum fluctuations, and effects of the chirality anomaly are expected to become stronger while the lifetime of the magnetic field decreases. It is difficult to quantitatively predict the trend of the CME amplitude as a function of the collision energy.
Therefore, a collision program with heavy isobar nuclei at the LHC would be valuable to experimentally answer the question whether or not the CME is observable at LHC energies.

In summary, after nearly two decades of intensive experimental and theoretical investigation, no definitive CME signal has yet been established, and a few percent level accuracy on  background contributions would be required for a firm conclusion.
Continued progress through higher-precision measurements, advanced correlation techniques, and complementary observables remains essential to fully uncover whether the CME, an imprint of QCD topology and quantum anomaly, manifests in the hot QCD medium created in heavy ion collisions.

\backmatter

\bmhead{Acknowledgments}

We thank many of our ALICE, CMS, and STAR collaborators for discussions and inputs. This work is supported by the U.S.~Department of Energy Office of Science, Office of Nuclear Physics under Awards No.~DE-SC0005131 (WL), No.~DE-SC0012910 (FW), and the National Natural Science Foundation of China under Grant No. 12322508 (QS).

\bibliography{./ref_notitle}
\end{document}